\documentclass[conference,compsoc]{IEEEtran}

\usepackage{enumerate}
\usepackage{enumitem}
\usepackage{graphicx}
\usepackage{epstopdf}
\usepackage{epsfig}
\usepackage{etoolbox}
\usepackage{tikz}
\usepackage{amsmath}
\usepackage{stmaryrd}
\usepackage{mathtools}
\usepackage{hhline}
\usepackage{setspace}
\usepackage{booktabs}
\usepackage{multicol}
\usepackage{multirow}
\usepackage{xcolor,colortbl}
\usepackage{array}
\usepackage{xspace}
\usepackage{pifont}
\usepackage{graphicx}
\usepackage{subcaption}
\usepackage[most]{tcolorbox}
\usepackage[htt]{hyphenat}
\usepackage{float}
\usepackage[flushleft]{threeparttable}
\usepackage[multiple]{footmisc}
\usepackage{adjustbox}
\usepackage{diagbox}
\usetikzlibrary{arrows}
\usepackage{wasysym}
\usepackage{longtable}

\usepackage{listings}
\usepackage{color}
\usepackage{enumitem}

\usepackage{fontawesome5}

\usepackage{hyperref}
\hypersetup{
  hidelinks,
	linkcolor = {darkpurple},
  breaklinks = true
}

\usepackage{cleveref}

\definecolor{dkgreen}{rgb}{0,0.6,0}
\definecolor{gray}{rgb}{0.5,0.5,0.5}
\definecolor{mauve}{rgb}{0.58,0,0.82}
\definecolor{Gray}{gray}{0.9}
\definecolor{persiangreen}{rgb}{0.0, 0.65, 0.58}
\definecolor{viridian}{rgb}{0.25, 0.51, 0.43}
\definecolor{ruddy}{rgb}{1.0, 0.0, 0.16}

\newcommand{\cmark}{{\color{black}\ding{51}}}%
\newcommand{\xmark}{{\color{ruddy}\ding{55}}}%

\newcommand{\system}{\texttt{AgentShield}\xspace}

\definecolor{borderblue}{HTML}{006EAF}
\definecolor{fillblue}{HTML}{1BA1E2}

\usepackage{pgfplots}
\pgfplotsset{compat=1.16}

\definecolor{lightgray}{rgb}{0.95, 0.95, 0.95}
\definecolor{darkgray}{rgb}{0.4, 0.4, 0.4}
\definecolor{purple}{rgb}{0.65, 0.12, 0.82}
\definecolor{editorOcher}{rgb}{1, 0.5, 0} 
\definecolor{editorGreen}{rgb}{0, 0.5, 0} 
\definecolor{orange}{rgb}{1,0.45,0.13}		
\definecolor{olive}{rgb}{0.17,0.59,0.20}
\definecolor{brown}{rgb}{0.69,0.31,0.31}
\definecolor{purple}{rgb}{0.38,0.18,0.81}
\definecolor{lightblue}{rgb}{0.1,0.57,0.7}
\definecolor{lightred}{rgb}{1,0.4,0.5}

\definecolor{darkpurple}{rgb}{0.502, 0.26, 0.747}
\definecolor{background}{HTML}{EEEEEE}
\definecolor{commentgreen}{RGB}{2,112,10}
\colorlet{numb}{magenta!60!black}
\definecolor{eminence}{RGB}{108,48,130}

\usepackage{xurl}
\usepackage{upquote}

\definecolor{eclipseStrings}{RGB}{42,0.0,255}
\definecolor{eclipseKeywords}{RGB}{127,0,85}
\colorlet{numb}{magenta!60!black}
\lstdefinelanguage{json}{
  keywordstyle=\color{blue}\bfseries,
  keywords={permissions, content\_scripts,manifest\_version,background,browser\_action},
  keywordstyle=\color{lightblue},
  identifierstyle=\color{black},
  sensitive=false,
  comment=[l]{//},
  morecomment=[s]{/*}{*/},
  commentstyle=\color{purple}\ttfamily,
  stringstyle=\color{red}\ttfamily,
  frame=single,
    literate=
     *{0}{{{\color{numb}0}}}{1}
      {1}{{{\color{numb}1}}}{1}
      {2}{{{\color{numb}2}}}{1}
      {3}{{{\color{numb}3}}}{1}
      {4}{{{\color{numb}4}}}{1}
      {5}{{{\color{numb}5}}}{1}
      {6}{{{\color{numb}6}}}{1}
      {7}{{{\color{numb}7}}}{1}
      {8}{{{\color{numb}8}}}{1}
      {9}{{{\color{numb}9}}}{1}
      {\{}{{{\color{delim}{\{}}}}{1}
      {\}}{{{\color{delim}{\}}}}}{1}
      {[}{{{\color{delim}{[}}}}{1}
      {]}{{{\color{delim}{]}}}}{1},
}

\definecolor{delim}{RGB}{20,105,176}
\definecolor{numb}{RGB}{106, 109, 32}
\definecolor{string}{rgb}{0.64,0.08,0.08}

\lstdefinelanguage{Swift}{
  keywords={typeof, new, true, false, catch, function, func, async, return, null, catch, switch, var, if, in, while, do, else, case, break, AVCaptureSession, AVCaptureDeviceInput, AVCaptureDevice, video, authorized, notDetermined},
  keywordstyle=\color{blue}\bfseries,
  ndkeywords={class, export, boolean, throw, implements, import, this},
  ndkeywordstyle=\color{darkgray}\bfseries,
  identifierstyle=\color{black},
  sensitive=false,
  comment=[l]{//},
  morecomment=[s]{/*}{*/},
  commentstyle=\color{purple}\ttfamily,
  frame=single,
  stringstyle=\color{red}\ttfamily,
  morestring=[b]',
  morestring=[b]"
}

\lstdefinelanguage{json}{
  keywordstyle=\color{blue}\bfseries,
  keywords={Camera, Data,Type, Purpose, Handling, User, Permission, API, Access, true, false, notdetermined},
  keywordstyle=\color{lightblue},
  identifierstyle=\color{black},
  sensitive=false,
  comment=[l]{//},
  morecomment=[s]{/*}{*/},
  commentstyle=\color{purple}\ttfamily,
  stringstyle=\color{red}\ttfamily,
  frame=single,
    literate=
     *{0}{{{\color{numb}0}}}{1}
      {1}{{{\color{numb}1}}}{1}
      {2}{{{\color{numb}2}}}{1}
      {3}{{{\color{numb}3}}}{1}
      {4}{{{\color{numb}4}}}{1}
      {5}{{{\color{numb}5}}}{1}
      {6}{{{\color{numb}6}}}{1}
      {7}{{{\color{numb}7}}}{1}
      {8}{{{\color{numb}8}}}{1}
      {9}{{{\color{numb}9}}}{1}
      {\{}{{{\color{delim}{\{}}}}{1}
      {\}}{{{\color{delim}{\}}}}}{1}
      {[}{{{\color{delim}{[}}}}{1}
      {]}{{{\color{delim}{]}}}}{1},
}

\begin{document}
\bstctlcite{BSTcontrol}

\title{\texttt{AgentShield:} Fortifying Web Applications Against\\AI Agents and Agentic Browsers}
\author{
\IEEEauthorblockN{Moiz Ali\textsuperscript{*\S}, Nafis Karim\textsuperscript{*\S}, Rownak Islam\textsuperscript{*}, Jyotirmay Chauhan\textsuperscript{*}, Dhruv Kuchhal\textsuperscript{\dag\ddag}, Jason Polakis\textsuperscript{*}}
\IEEEauthorblockA{\textit{\textsuperscript{*}University of Illinois Chicago, \textsuperscript{\dag}Amazon.com, Inc.}\\
\{sali269, nkari22, misla84, jchauh3, polakis\}@uic.edu, dkuchhal@amazon.com}
}
\maketitle
\begingroup\renewcommand\thefootnote{\S}
\footnotetext{These authors contributed equally to this work.}
\endgroup
\begingroup\renewcommand\thefootnote{\ddag}
\footnotetext{This work is unrelated to the author's position at Amazon.com, Inc.}
\endgroup

\begin{abstract}


LLM-powered browser agents are transforming web automation by autonomously
interacting with websites and executing complex workflows. However, the
effectiveness of existing anti-bot defenses against these systems remains
unclear. We present a systematic empirical evaluation of modern agentic
browsers against widely deployed anti-bot mechanisms, characterizing their
capabilities, limitations, and behavioral differences from humans. Our
results show that many defenses designed for traditional automation
struggle to reliably detect autonomous AI-driven agents.  Accordingly, we
introduce \system, a behavioral fingerprinting framework that targets the
Perception–Reasoning–Action (PRA) loop underlying agentic user clients. By
leveraging dynamic web forms that progressively reveal content and capture
fine-grained interaction telemetry, \system exposes behavioral artifacts
that enable differentiation between human users and autonomous agents. We
further demonstrate Extinction-induced Policy Drift (EPD), a novel attack
in which repeated obstruction of task completion causes agents to relax
their safeguards and engage in unsafe behaviors. Finally, we present
defensive prompt injection techniques that exploit agent-specific
weaknesses to mitigate emerging forms of autonomous web abuse that rely on
content extraction. Our results highlight both the shortcomings of current
web defenses against agentic user clients and the feasibility of practical
detection mechanisms tailored to this new generation of web automation.

\end{abstract}

\section{Introduction}\label{sec:intro}

Fundamentally, cyberattacks rely on automation to achieve scale. Credential
stuffing~\cite{thomas2019protecting,wang2020detecting}, fake-account
creation~\cite{yuan2019detecting,drakonakis2020cookie},
spam~\cite{stone2011underground,grier2010spam,simoiu2020targeted}, and scraping
depend on emulating user activity across sites. Historically, attackers
have used scripted bots built on frameworks (e.g., Selenium, Puppeteer),
often through headless browsers~\cite{li2021good}. Services have responded with
anti-bot defenses~\cite{amin2020web}, including
CAPTCHAs~\cite{sivakorn2016robot,searles2023captcha}, browser
fingerprinting~\cite{laperdrix2020browser,wu2023him}, and IP
reputation~\cite{hao2009detecting}. These defenses exploit artifacts of
traditional automation in browser APIs~\cite{vastel2020fp}, JS
execution, timing and resource access~\cite{fontana2025no}, and network
behavior~\cite{amin2020web}.

The threat landscape is now shifting as LLMs enable increasingly capable
autonomous agents. Cloudflare reports 18\% year-over-year growth in AI-crawler
traffic, with GPTBot's share rising from 5\% to 30\%, while only $\sim$14\% of top
sites publish \texttt{robots.txt} rules targeting AI
agents~\cite{cloudflare_crawlers_2025}. Recent LLMs demonstrate sophisticated
reasoning, planning, and tool
use~\cite{Singhal2025-gx,chen-llmcodeeval,gpt4bar2024,llm-quantitative-reasoning},
enabling agentic systems to autonomously browse the web, fill forms, execute
transactions, and interact with online services. Products such as Perplexity
Comet and OpenAI's browser agents are bringing these capabilities to consumers,
while reports already document their use in phishing and large-scale
scams~\cite{openai-malicious-uses,gressel2026lovelieslanguagemodels,badhe2025scamagents}.
Recent studies have also shown that agentic browsers can be
manipulated by untrusted web content, including hidden instructions that drive
account takeover, fraudulent purchases, and credential
submission~\cite{brave_comet_injection_2025,guardio_scamlexity_2025}.

Thus, agentic browsers and AI-powered web automation tools pose a major new
challenge to securing the web ecosystem, as existing anti-bot assumptions based
on static browser fingerprints or deterministic automation patterns may no
longer hold~\cite{cloudflare_perplexity_stealth_2025,chatgpt_agent_captcha_2025}.
%
As agentic browsing becomes more widely deployed, it will become increasingly
important for web services to reliably identify, regulate, or selectively
restrict such clients in order to preserve platform integrity, defend
monetization models, and prevent widespread abuse.  Accordingly, the research
community has begun to investigate the security implications of agentic web
systems. Recent studies demonstrate indirect prompt injection attacks against browser
agents~\cite{greshake2023indirect, kumar2024refusal, brave_comet_injection_2025}, social engineering attacks targeting autonomous
workflows~\cite{darkpatterns2025}, and fingerprinting techniques
that distinguish AI-driven browsers from human
users~\cite{ayzenshteyn2025cloak}.
However, despite increasing concern regarding the abuse potential of agentic
clients, we currently lack a systematic evaluation of emerging agentic systems
against existing anti-bot defenses. 

In this paper, we fill that gap by presenting a systematic empirical
evaluation of agentic user clients against widely deployed anti-bot defenses.
We study the capabilities of agentic browsers and automation
frameworks when interacting with widely-available protection systems, including
Google's reCAPTCHA, and Cloudflare's Turnstile.
Through extensive experiments
across multiple browsing tasks, we characterize the
extent to which current agentic systems bypass, evade, or falter against
existing bot mitigation techniques. Our findings provide one of the first
large-scale empirical perspectives on how agentic clients
meaningfully alter the current balance between web task automation and anti-bot
defenses. Building on these observations, we present \system, a novel
behavioral fingerprinting framework that detects agentic clients by
exploiting inherent limitations and behavioral artifacts in how autonomous
agents navigate and interact with websites. Through extensive experiments, we
demonstrate how \system detects modern agentic clients using specially
crafted dynamic forms that capture granular information during form
interactions and target the Perception$\rightarrow$Reasoning$\rightarrow$Action (PRA)
loop~\cite{pra-loop} that 
drives agentic clients. Modern web apps can readily
deploy our approach to fortify critical processes (e.g.,~account creation).

Moreover, while prior work has demonstrated prompt injection techniques that
can reveal the use of agentic clients (or exploit them), modern agents attempt
to counteract such techniques through explicit \emph{safeguards}~\cite{openai-agent-safeguards,browsesafe-perplexity, prompt-injection-anthropic}.
To that end, we demonstrate an attack we call
Extinction-induced Policy Drift (EPD),\footnote{In psychology,
\emph{Extinction Burst}~\cite{katz2021extinction} describes cases where a person's behavior temporarily
becomes increasingly erratic when a previously rewarded behavior suddenly stops
producing rewards.} wherein interaction with state-changing web elements (e.g.,
a form submission button) does not result in any changes, leading to a failure
mode where repeated obstruction of task completion causes an agentic browser to
relax constraints and pursue unsafe actions in
search of reward completion (i.e., the agent overrides its internal
safeguards). Since our goal is to provide a holistic defense against diverse agentic workflows, we also 
focus on passive content extraction which may not expose the aforementioned behavioral signals.
To that end, we demonstrate a set of novel defensive prompt
injection techniques targeting two emerging abuse classes enabled by agentic
systems: large-scale e-commerce scraping operations that replicate product
pages, and automated news aggregation systems that summarize or repost
copyrighted articles on third-party platforms~\cite{hiltbrand2024guarding,grimmelmann2026ai,nyt-lawsuit-perplexity}.
Overall, our research highlights both the growing risks posed by autonomous
browsing agents and the feasibility of developing practical defenses tailored
to this new generation of automation. In summary, we make the following 
contributions:

\begin{itemize}
\item We present a systematic evaluation of agentic clients' capabilities against prevalent
anti-bot mechanisms, and demonstrate the risk they pose to web services.

\item We identify key behavioral characteristics of agentic clients, and propose
a detection mechanism that targets critical browsing workflows (e.g., account creation).

\item We present a novel technique that drives agentic clients to ignore internal safeguards
and perform prohibited actions, which in turn reveals their presence. We also propose two
defensive prompt injection attacks for protecting web content from illicit
scraping. We present extensive experimental results that demonstrate \system's effectiveness and robustness, 
even against adaptive attackers with knowledge of the signals targeted by our system.

\item We implement \system as installable NPM packages and will publicly release all of our code, agent prompts, and data to facilitate further research~\cite{agentshield_repo}.
\end{itemize}

\section{Background}
\label{sec:background}

This section provides background on modern AI user clients and their capabilities. We will use \emph{AI user clients} to refer to the various types 
of agent-driven tools. 


\begin{table}[!t]
  \centering
  \caption{AI clients evaluated in this study. \textit{Execution} indicates
    whether the client's request originates from the user's machine, a
    cloud-hosted environment, or both.}
  \resizebox{\linewidth}{!}{%
 \begin{tabular}{l|l|l|l|l} 
  \toprule
  \textbf{Client} & \textbf{Provider} & \textbf{Model} & \textbf{Execution} & \textbf{Perception} \\
  \midrule
  \rowcolor{lightgray}
  \multicolumn{5}{l}{\textit{Autonomous Web Agents}} \\
  \midrule
  Perplexity Comet & Perplexity  & Default  & Local-Cloud & Hybrid \\
  ChatGPT Atlas   & OpenAI      & GPT-5.5  & Local-Cloud & Hybrid \\
  Claude for Chrome & Anthropic       & Opus 4.8  & Local & Hybrid \\
  Amazon Nova Act    & Amazon      & Nova 2 Lite  & Cloud &  Vision \\
  Director.ai     & Browserbase & Default  & Cloud & Hybrid \\
  Manus           & Meta      & 1.6 Lite  & Cloud & Hybrid \\
  Devtools MCP (Claude) & Google & Opus 4.8 & Local & Hybrid \\
  \midrule
  \rowcolor{lightgray}
  \multicolumn{5}{l}{\textit{Browser Assistants}} \\
  \midrule
  Gemini in Chrome & Google     & Gemini 3.5 Flash    & Local & Hybrid \\
  Edge Copilot    & Microsoft   & Default       & Local & DOM \\
  Brave Leo       & Brave       & Default     & Local & DOM \\
  \midrule
  \rowcolor{lightgray}
  \multicolumn{5}{l}{\textit{Web LLMs}} \\
  \midrule
  ChatGPT         & OpenAI      & GPT-5.5 & Cloud & HTML \\
  Claude          & Anthropic   & Sonnet 4.8  & Cloud & HTML \\
  Gemini          & Google      & 3.1 Pro         & Cloud & HTML \\
  \midrule
  \rowcolor{lightgray}
  \multicolumn{5}{l}{\textit{Web Scrapers}} \\
  \midrule
  Crawl4AI        & Open source & Gemini 3 Flash & Local & DOM \\
  \bottomrule
  \end{tabular}
  }
\label{tab:agents-tested} 
  \end{table}

\textbf{AI user clients.} An AI client is a program that uses a large language
model (LLM) to read or interact with the web on a user's behalf. The user gives
it a task in natural language, and the tool then visits the relevant web pages,
reads their content, and autonomously decides what to do next to complete the
task, without following a fixed script. This makes AI clients fundamentally
different from traditional bots. A traditional bot typically follows hard-coded
selectors or heuristics to complete a pre-determined series of actions, and can
easily break when a page changes. In contrast, AI user clients are designed to
adapt according to the specifics of the website they are interacting with; they
render JavaScript, interpret page layout and content semantics, and can follow
instructions written for humans.



\textbf{Taxonomy.}
We categorize AI user clients based on two
classes of properties that are critical from the perspective of a web server
receiving a request.
\begin{enumerate}[label=(\roman*)]
\item \emph{Access mechanism.} Here we focus on whether
the client drives a full browser, issues raw HTTP fetches, reads a tab the
user already has open, or converts pages into LLM-ready formats.

\item \emph{Execution environment.} Here we focus on whether the automation
workflow is running locally on the user's machine or in a cloud-hosted
environment. 
\end{enumerate}

These two properties determine the signals a server can collect: whether the
client runs JavaScript, whether it triggers mouse and keyboard events, what its
TLS and HTTP fingerprints look like, and whether the request belongs to an
established, legitimate user session. We use these properties to
assign the AI user clients we include in our study to four different categories,
as shown in~\Cref{tab:agents-tested}.

\emph{Autonomous Web Agents} are goal-driven systems that operate a full
browser through a Perception $\rightarrow$ Reasoning $\rightarrow$ Action loop.
At each step, the browser state is conveyed to an LLM, which reasons and emits
a tool call executed by automation frameworks (e.g.,
Playwright~\cite{playwright}, Chrome DevTools Protocol
(CDP)~\cite{chrome_devtools_protocol}). Agents use DOM-based, vision-based, or
hybrid perception~\cite{he-etal-2024-webvoyager}: DOM-based agents select
elements from the DOM or accessibility tree; vision-based act on
annotated screenshots; hybrid agents combine both, typically falling back
to vision when DOM information is insufficient. Agents may also run locally or
in the cloud, creating substantial differences at the network level.

\emph{Browser Assistants} are extensions or sidebars that read the page the
user already has open and pass its rendered content to an LLM, inheriting the
user's session. They do not navigate on their own. Gemini in Chrome, Edge
Copilot, and Brave Leo fall into this category.

\emph{Web LLMs} fetch pages over HTTP and parse the raw HTML directly,
without executing JavaScript, before passing the text to an LLM. ChatGPT,
Claude, and Gemini operate this way in their web-search mode.

\emph{Web Scrapers} extract page content into LLM-ready formats (e.g.,
Markdown, structured text, or screenshots) before passing it to a model.
We use Crawl4AI to represent this category; unlike the others,
it functions as a pipeline component rather than an interactive client.

\textbf{Anti-bot defenses.} Modern web services employ a wide range of mechanisms
to separate humans from automated clients. These defenses fall into three broad
categories.

\emph{Network-level} defenses examine the request itself, including blocking
specific IP ranges, applying rate limiting, and fingerprinting the TLS handshake
to recognize scripted clients that cannot replicate a real
browser's TLS signature. 
JA3 and its successor JA4~\cite{a2023_ja4} are the
prevalent TLS fingerprints: both hash fields from the client's
\texttt{ClientHello} message.
Since browsers and HTTP libraries
populate these fields differently, the fingerprints can differ across clients.
Commercial tools like 
Cloudflare's AI Crawl Control~\cite{a2026_aicrawlcontrol} and AI Bot 
Blocking~\cite{a2025_aibotblock} combine network and other signals (e.g., IP
reputation and self-declared identity) into a single service that decides
whether to block a request. 

\emph{Environment validation} tests clients at the browser layer using
CAPTCHAs, proof-of-work, browser-API challenges (e.g., Cloudflare Turnstile),
and device fingerprinting (e.g., FingerprintJS) to verify an interactive
browser.

\emph{Identity and signature-based} defenses instead identify clients through
declared or cryptographically verifiable identities. Traditional crawlers such
as Googlebot expose User-Agents, IP ranges, and \texttt{robots.txt} policies
that sites can use to allow, throttle, or block traffic. More recently,
providers such as Cloudflare offer verified-bot programs in which operators
register once for recognition across participating
sites~\cite{a2025_aibotblock}. Because User-Agents and IPs are easily forged,
newer approaches bind requests to cryptographic identities. For example,
Cloudflare's Web Bot Auth~\cite{a2026_webbotauth} uses HTTP Message
Signatures~\cite{sporny_2024_http}, allowing sites to verify requests against
an operator's published public key.

\emph{Challenges presented by AI user clients.} AI clients break many of the
assumptions behind these defenses. Autonomous agents drive real browsers, and
agents that run a browser locally on the user's machine issue requests from the
user's residential IP and inherit the user's session cookies. Browser
assistants make no requests of their own; they simply read the page the user has
already loaded. Moreover, identity-based defenses rely on the honor system:
declared AI User-Agents and signed-agent attestations only work for clients that
choose to identify themselves, so unsigned agents pass through unchecked. As AI
user clients proliferate, we must understand how they fare against prevalent
anti-bot defenses and whether new mechanisms can be tailored to detect
them.  Accordingly, in \Cref{subsec:current-defenses-eval} we present
a systematic evaluation that explores the effectiveness and limitations of
prevalent defenses against modern AI clients.


\textbf{Threat Model.} We adopt the perspective of a website that wishes to
detect or control how AI clients interact with its content. The defender
controls the web server, but cannot modify the AI client and cannot assume
cooperation from it. The adversary leverages an off-the-shelf AI user client,
which may run locally or in the cloud, may or may not declare itself, and acts
under instructions from a human operator whose intent is not known. While
adversaries may in practice run local LLMs without safeguards, our study
focuses on how popular off-the-shelf AI user clients behave across scenarios,
and the effect of their built-in safeguards.

\section{Methodology}
\label{sec:methodology}

This section details our analysis pipeline for agentic user clients,
which combines prevalent existing tools and
services with a series of novel mechanisms we have developed.


\subsection{Current Defenses}
\label{subsec:current-defenses}

We test three classes of existing defenses against our AI client set:
\emph{network signals}, \emph{environment-validation challenges}, and
\emph{self-declared identity}. 
\Cref{tab:defense-classes} maps
each defense to its class. Most fall in one class, though Cloudflare's Crawl
Control and AI Bot Blocking spans both network and self-declared identity
signals.

\begin{table}[t]
\centering
\caption{Existing defense categorization.} \label{tab:defense-classes}
\resizebox{\linewidth}{!}{%
\begin{tabular}{l ccc}
\toprule
& \shortstack{\textbf{Network}\\\textbf{Signals}}
& \shortstack{\textbf{Environment}\\\textbf{Validation}}
& \shortstack{\textbf{Self-}\\\textbf{Declared}} \\
\midrule
\rowcolor{lightgray}
IP Filtering             & \CIRCLE & \Circle & \Circle \\
TLS Fingerprint (JA4)    & \CIRCLE & \Circle & \Circle \\
\rowcolor{lightgray}
Dual-Request Pattern     & \CIRCLE & \Circle & \Circle \\
Cloudflare Crawl Control & \CIRCLE & \Circle & \CIRCLE \\
\rowcolor{lightgray}
reCAPTCHA v3             & \Circle & \CIRCLE & \Circle \\
Cloudflare Turnstile     & \Circle & \CIRCLE & \Circle \\
\rowcolor{lightgray}
User-Agent               & \Circle & \Circle & \CIRCLE \\
Request Signing          & \Circle & \Circle & \CIRCLE \\
\bottomrule
\end{tabular}
}
\end{table}


\textbf{IP Filtering.} The server records the source IP and classifies it as
data-center or residential through an IP-reputation provider. This signal
targets agents that use cloud infrastructure. Agents running on the user's
local machine or behind residential proxies are not detected at this layer.

\textbf{TLS Fingerprint (JA4).} A Zeek SSL logger~\cite{a2020_zeek} deployed on
the server records the JA4 fingerprint of every TLS handshake. 
Real browsers produce a standard Chrome fingerprint, whereas raw HTTP libraries
and atypical TLS stacks produce non-standard fingerprints that we flag.

\textbf{Dual-Request.} Some local-browser agents issue a separate
server-side request from their cloud backend alongside the local browser
session. 
We flag any pair of requests to the same page from two different IPs within a
short time window. A normal human visit does not produce this pattern.

\textbf{Cloudflare Crawl Control and AI Bot Blocking.} We route the test site
through Cloudflare with AI Crawl Control~\cite{a2026_aicrawlcontrol} and AI Bot
Blocking~\cite{a2025_aibotblock} enabled in managed mode. Cloudflare uses
IP reputation, declared User-Agent, and its own signed-agent registry to decide
whether to block.

\textbf{reCAPTCHA v3.} We deploy Google's reCAPTCHA
v3~\cite{google_2019_recaptcha} on a test page. The challenge runs silently in
the background, scoring the client on behavioral signals and cross-site
browsing history, and returns a bot-likelihood score between 0 and 1. We treat
a score below the default threshold of 0.5 as a block verdict.

\textbf{Cloudflare Turnstile.} We deploy Turnstile on a similar test page.
Turnstile issues a combination of proof-of-work, proof-of-space, and
browser-API challenges that target headless browsers and non-standard
environments. Each agent receives either a pass or a failure verdict on each
session.


\textbf{User-Agent.} The server logs the UA string on every request. An
agent that declares itself through a UA suffix (e.g.,
\texttt{Perplexity-User/1.0}) 
self-identifies at this layer.

\textbf{Request Signing.} The server inspects each request for the
\texttt{Signature-Agent}, \texttt{Signature-Input}, and \texttt{Signature}
headers defined by RFC~9421~\cite{sporny_2024_http} and Cloudflare's Web Bot
Auth draft~\cite{a2026_webbotauth}. A signed request cryptographically attests
to the agent's identity.

\subsection{Dynamic Forms}
\label{subsec:dynamic-forms}

Forms constitute a foundational component of consequential web
interactions (e.g., account creation, sign-in, checkout, job applications, and
onboarding workflows) and represent a persistent friction point for users.
This friction is precisely the capability that agentic browsers advertise as a
primary value proposition. Perplexity's Comet markets form-filling as a core
feature of its sidebar AI assistant~\cite{a2025_comet_quickstart}; Manus
positions its Browser Operator around eliminating manual clicking and form
completion~\cite{a2025_manus_browser_operator}; OpenAI similarly frames ChatGPT
Atlas's agent mode as automating multi-step browser workflows, exemplified by
converting a recipe into a grocery order through tab opening, cart population,
and order placement~\cite{a2025_introducing_atlas}. Given that forms gate
substantial portions of the authenticated web, and browser vendors explicitly
promote form automation as a distinguishing capability, we employ dynamic forms
as our initial investigation into agentic behavior.

Dynamic forms are web forms in which fields are concealed from the user and
revealed sequentially upon completion of the preceding field. 
We implement concealment via CSS
properties (\texttt{opacity: 0} and zero dimensions), so that fields remain
present in the DOM but invisible until triggered. This design necessitates that
any agentic client observe the page after each field reveal before proceeding,
thereby making the latency of the agent's Perception $\rightarrow$Reasoning$\rightarrow$Act 
loop directly measurable and comparable across humans,
browser autofill, and AI agents.

\textbf{Form Design.} 
Our form is a signup page containing 
22 conditionally-revealed
fields that span the input types commonly found in real-world forms: text and
telephone inputs, email, date pickers, single-select dropdowns, radio buttons,
checkboxes, and a free-form textarea. The reveal sequence is branching rather
than linear: selecting a country exposes a country-specific sub-field 
(e.g., US~$\rightarrow$~state~$\rightarrow$~city, Canada~$\rightarrow$~province);
choosing a volunteer role reveals a role-specific panel (field research, public
education, fundraising, or remote support); and certain selections (full-time
availability, social-media or friend-referral as the source) trigger additional
follow-up fields. These branching structures enable a single session to
exercise multiple input types and yield sufficient events for stable statistics
across each metric.
A JavaScript handler attached to each preceding field reveals the subsequent
field upon event completion. Text fields trigger on the \texttt{input} event
once a minimum character threshold is satisfied; selects, radios, and
checkboxes trigger on the \texttt{change} event. Upon reveal, the handler
appends a \texttt{visible} class to the target field's container, which the
\texttt{MutationObserver} detects as the reveal signal. This mechanism ensures
that field visibility transitions are observable events, enabling precise
timestamping of each reveal.

\textbf{Behavioral Metrics.} The form's design is motivated by four behavioral
metrics that collectively distinguish agentic from human interaction patterns.
(i)~\textit{Dwell Time} (duration between \texttt{keydown} and
\texttt{keyup} for each keystroke, reported as per-session mean and standard
deviation) is measurable because the form's sequential reveal isolates
individual keystrokes within discrete interaction episodes. Auto-repeat presses
from held keys are excluded to prevent count inflation. (ii)~\textit{Field
Reaction Time} (interval between a field's reveal and the client's first
interaction with it) is directly enabled by the CSS concealment mechanism: the
\texttt{MutationObserver} provides reveal timestamps (triggering on the
concealed-to-visible transition), while the field's \texttt{input} and
\texttt{change} events provide interaction timestamps. 
This interval captures the end-to-end
Perception$\rightarrow$Reasoning$\rightarrow$Act loop latency, which for agents is
dominated by LLM inference time rather than typing or navigation speed.
(iii)~\textit{Keypress Count} (total number of \texttt{keydown} events per
session) reveals clients that inject or paste values rather than typing them,
as these produce near-zero counts. (iv)~\textit{Mouse Movement Count} (total number of
\texttt{mousemove} samples per session at a 40ms sampling rate) is
meaningful because the form's multi-field layout requires pointer navigation
across spatially distinct targets. 
Because browser-automation frameworks typically
emit pointer events only at click targets rather than along a continuous
trajectory, agents produce near-zero counts, whereas humans generate continuous
traces across the form. Human typing exhibits high variance across the timing
metrics due to finger transitions, corrections, and cognitive pauses; agents
produce mechanically consistent values with near-zero spread.

Client-side JavaScript injected into the form page captures browser events at
fine granularity. The DOM Event API~\cite{a2025_dom} records mouse activity,
keystrokes, and form-level input and change events; the MutationObserver Web
API~\cite{a2026_mutationobserver} records field-reveal timestamps by triggering
on every DOM mutation that transitions a concealed field into a visible state.
Mouse clicks (\texttt{mousedown}) are recorded individually, and
\texttt{keydown}/\texttt{keyup} timestamps are captured per keystroke. Every
event is timestamped, enabling full session-timeline reconstruction.

\textbf{Deployment.} The dynamic form has two variants:
\begin{enumerate}
    \item \textit{Standard variant}: includes dropdown fields. However, we found 
    that certain agents cannot navigate dropdown menus reliably.
    \item \textit{Simplified variant}: excludes dropdown fields to establish a 
    common baseline across all clients.
\end{enumerate}

\subsection{Content Modification}
\label{subsec:content-modification}

While our dynamic forms detect agentic behavior through interaction patterns,
certain agentic workflows may not include such interactions (e.g., an agentic scraper). As such, our goal of providing 
a holistic defense against modern agentic clients necessitates
a complementary technique: content-based poisoning of pages
consumed by agents. Unlike traditional scrapers, agentic clients interpret page
content through an LLM, making them susceptible to semantic manipulation. Even
when perimeter defenses are bypassed, poisoned content can corrupt agent
outputs by inducing compliance with hidden instructions or relaying fabricated
information. This vulnerability is intrinsic to the PRA loop, whose reasoning
module processes DOM text regardless of whether it is intended for human
consumption.

\textbf{Testbed Configuration.} We deploy a WordPress-based test site combining
an e-commerce store and news blog, with injections placed in (i) news and blog
articles covering food, health, and environmental topics, and (ii) product
listings, including descriptions, prices, and attributes.

\textbf{Hidden Prompt Taxonomy.} We embed prompts in the DOM that are invisible
to humans but accessible to agents. We evaluate five manipulation strategies,
using one fixed prompt per category across all clients and runs:
\emph{Defensive Prompt} claims that content is proprietary and extraction
violates site policy. Success occurs when the agent refuses, halts extraction,
or treats the notice as a reason for caution.
\emph{Hidden Promotion} instructs the agent to include a fabricated marketing
offer absent from the visible page; success occurs when the offer appears in
its response.
\emph{Opposite Information} instructs the agent to replace a visible claim
with a contradictory statement; success occurs when the contradiction is
presented as factual.
\emph{Additional Information} asks the agent to append an unsupported but
plausible detail; success occurs when the injected detail appears in its
output.
\emph{Abusive Prompt} injects a threatening or alarming warning intended to
derail the response; success occurs when the agent reproduces or treats the
warning as a genuine page-level safety concern.


Of these five categories, only \emph{Defensive Prompt} has a direct analogue in prior work.
WebCloak~\cite{li2026webcloak} injected anti-extraction notices that invoke site
policy or trigger LLM safety alignment, but did so to protect \emph{images}
from scraping rather than to manipulate an agent's textual output. Greshake et
al.~\cite{greshake2023indirect} introduced indirect prompt injection as a way to
embed hidden instructions in web content, including factual manipulation that
overlaps with our \emph{Opposite Information} category, but they showed that
injection is possible without measuring each manipulation goal across deployed
agents. The remaining three categories are novel types of
page-level injections against agentic web clients.

\textbf{Concealment Method.} A hidden prompt reaches an agent only if it
resides in the content layer that the agent perceives, and the clients in our
study do not share one: text-parsing clients read the DOM, the raw HTML, or the
accessibility tree, whereas vision-based clients reason exclusively over a
rendered screenshot. No single technique covers all, so we employ two
complementary approaches. 

\emph{DOM-layer concealment.} We conceal hidden
prompts using a CSS class that makes the injected text invisible on screen
while keeping it in the DOM. The class sets font size and line height to zero,
forces the color to transparent, and uses absolute positioning with a one-pixel
bounding box to push the element out of the visible layout. Nothing appears on
screen and no whitespace is introduced, but the text sits in the document tree
and is readable by any client that processes raw HTML or the DOM. We apply this
\texttt{.zero-text-injection} class across all five prompt categories in
the test pages (see \Cref{fig:css-injection} in the Appendix). We use this approach instead of
\texttt{display:none} or \texttt{visibility:hidden} because both cause some
parsers and browser extensions to drop the element from their content
representation. Keeping the text in the document flow means agents working from
rendered content, raw HTML, or an accessibility tree all encounter the
injection. The \texttt{user-select: text} property is intentionally preserved:
it ensures the text remains selectable by clients that build their context from
user-selectable content rather than raw DOM parsing. The exception is agents
that operate purely from screenshots: for those, anything not visible in the
rendered image does not exist, regardless of what the DOM contains.

\emph{Visual-layer concealment.} To target the screenshot-based agents that
cannot be affected by the DOM technique, we introduce a complementary
concealment technique that places the injected text in the visual rendering
layer, drawn with the script in \Cref{fig:canvas-injection} (Appendix).  
In the primary configuration, the visible article text remains in the
DOM as normal page content, and only the injected prompt is moved into an HTML
\texttt{<canvas>} element, drawn in a low-opacity gray that is imperceptible to
humans.  We empirically tuned the rendering parameters to the threshold at
which the text remains imperceptible to humans yet recoverable by OCR: a fill
of \texttt{rgba(180,180,180,0.15)} at a 15\,px sans-serif size. Below this
opacity or font size, the agents that otherwise act on the visual-layer prompt
fail to recover it, indicating that the channel operates within a bounded
window where the text is dark and large enough for OCR while still remaining
imperceptible to humans (see Figure~\ref{fig:ocr-page}, Appendix~\ref{app:prompt-content-modification}).  
Since canvas content exists only as rendered pixels, the injected prompt is absent
from the DOM, the raw HTML, and the accessibility tree: a client that parses
the document tree sees an empty canvas element and recovers nothing, whereas a
client that rasterizes the page and applies OCR will recover the hidden prompt.
This inverts the agent coverage achieved by the DOM technique. 
For clients that recover nothing in this configuration,
we additionally render the entire article, including the visible body text,
onto the canvas alongside the hidden prompt, so that the full page content
exists only as pixels; this isolates whether a client rasterizes the page at
all, since a client that reads the page visually would recover the visible
canvas text even if the low-opacity prompt remained below its threshold.


\textbf{Repetition Variant.} To assess whether compliance scales with
repetition, we embed the same defensive prompt $N$ times, with $N \in \{3,
5, 7\}$, throughout the same page.
Each repeated instance is inserted at a
distinct location in the document using the same concealment method. 

\subsection{Extinction-Induced Policy Drift}
\label{subsec:delayed-response}

Loading screens and deliberate delays are a routine pattern on the web, but for
an autonomous agent, a stalled page is a decision trap. When an agent clicks a
state-changing element (e.g., a checkout or submit button) and the page
produces no observable change, its primary path to task completion is suddenly
withheld. 
A human reads this as a loading state and waits; an agent with an unmet
objective must instead decide whether to wait, retry, or seek another path.  We
show that under continuous obstruction, agents exhibit what we call
\emph{Extinction-Induced Policy Drift} (EPD): mirroring the \emph{extinction
burst} in behavioral psychology (in which a person's behavior becomes erratic
once the reward stops). The agent's actions become erratic and it relaxes its own
constraints in search of reward completion, eventually overriding its own
internal safeguards. We implement EPD by embedding a concealed instruction inside a deliberately
stalled page. An agent acting on this instruction, while having ignored it under normal conditions, 
is our signal that an agent has relaxed its safeguards.

\textbf{Setup.} We built 
a custom e-commerce platform where upon clicking on the secure checkout button
the product page ``freezes'' for 120s
before automatically redirecting to the payment form.
The product page includes a hidden prompt in the form of a
zero-dimension \texttt{div}, concealed with the same CSS class used in
\S\ref{subsec:content-modification}.
The instruction is framed as a site-generated fallback
mechanism, referencing the goal of the agent to purchase the product and
presenting an alternative URL as the recovery path
(see Appendix~\ref{app:delayed-response}).

\begin{table}[!t]
\centering
\caption{Network signals: source \textbf{IP} address type,
Cloudflare's AI Crawl Control (\textbf{CC}) + AI Blocking (\textbf{AIB}) verdict, and
presence of the JA4 dual-request pattern (\textbf{JA4-DR}).\\ 
\cmark = detected by signal; \xmark = not detected;}
\footnotesize
\begin{tabular}{l|lcc}
\toprule
\textbf{Agent} & \textbf{IP Type} & \textbf{CC + AIB} & \textbf{JA4-DR} \\
\midrule
\rowcolor{lightgray}
Comet               & Both                     & \xmark & \cmark \\
Nova                & Data Center              & \xmark & \xmark  \\
\rowcolor{lightgray}
Manus               & Data Center              & \cmark     & \xmark  \\
Director.ai         & Residential              & \xmark & \xmark  \\
\rowcolor{lightgray}
Atlas               & Both                     & \xmark & \cmark \\
Claude for Chrome   & Residential              & \xmark & \xmark  \\
\rowcolor{lightgray}
Chrome Devtools MCP & Residential              & \xmark & \xmark  \\
\midrule
ChatGPT             & Data Center              & \cmark     & \xmark  \\
\rowcolor{lightgray}
Claude              & Data Center              & \cmark     & \xmark  \\
Gemini              & Data Center              & \xmark & \xmark  \\
\midrule
\rowcolor{lightgray}
Crawl4AI            & Residential              & \xmark & \xmark  \\
\bottomrule
\end{tabular}
\label{tab:existing-defenses}
\end{table}

\begin{table}[!t]
\centering
    \caption{Environment-validation defenses.}
    \footnotesize
\begin{threeparttable}
\begin{tabular}{l|c|c}
\toprule
\textbf{Agent} & \textbf{reCAPTCHA v3} & \textbf{Turnstile} \\
\midrule
\rowcolor{lightgray}
Comet    & \xmark     & \xmark \\
Nova     & N/A\tnote{1}        & \cmark \\
\rowcolor{lightgray}
Manus    & \cmark & \cmark \\
Director.ai  & \cmark & \xmark \\
\rowcolor{lightgray}
Atlas    & \cmark & \xmark \\
Claude for Chrome   & \xmark     & \xmark \\
\rowcolor{lightgray}
Chrome Devtools MCP  & \cmark  & \cmark \\
\bottomrule
\end{tabular}
\begin{tablenotes}
\footnotesize
\item[1] Nova did not perform the task with reCAPTCHA v3 present.
\end{tablenotes}
\end{threeparttable}
\label{tab:env-defenses}
\end{table}

\section{Experimental Evaluation} 
\label{sec:evaluation}

Here we present our experimental evaluation of agentic user clients'
capabilities and limitations. Certain experiments were conducted only with a
subset of the agents from \S\ref{sec:background}. All of our agent prompts
can be found here~\cite{agentshield_repo}.

\subsection{Current Defenses}
\label{subsec:current-defenses-eval}

\textbf{Network signals.} 
We evaluate each class of existing defenses against the agentic user clients from \Cref{tab:agents-tested}, and demonstrate
how every class of signal can be evaded. 
We exclude
Browser Assistants since they issue no requests of their own and only read the page
already loaded in the browser. 
As shown in \Cref{tab:existing-defenses}, IP filtering catches cloud-hosted agents only when they expose
a data-center IP rather than routing through a proxy. 
Agents that run on cloud infrastructure have IP addresses
from subnets assigned to data centers. Crawl4AI and Director.ai originate from residential hosts
since we ran Crawl4AI locally, and Director.ai uses
Browserbase, which routes through residential proxies.
Naturally, the agents that run the browser locally on the
user's machine 
look indistinguishable from a human visitor at this layer. Comet and Atlas are marked with
``Both'' because their cloud backend issues a separate request, which we
capture under \textit{Dual Req}.
Of the data-center agents Cloudflare could
plausibly flag, it only blocks three.
Since Cloudflare designed its managed AI controls around known crawler
user-agents and signed agent identities, this is insufficient for identifying
and blocking all agents.  Finally, we note that Comet and Atlas trigger two
requests to the test page from two different IP addresses: one from the user's
residential browser, and one from the provider's backend.  Notably,
Cloudflare's defenses do not block either of these agents.

\textbf{Environment validation.} 
Since reCAPTCHA requires an action to be completed to provide a score,
we only test the 7 web agents.
\Cref{tab:env-defenses} reports the reCAPTCHA v3 and Turnstile
verdicts per agent.
While reCAPTCHA is 
able to detect several of them, Comet and
Claude for Chrome are able to pass it. We note that the same agents that evade the
network layer defenses by running inside the user's browser also
pass the active challenges.
Adding CAPTCHAs to a site that already
deploys network-layer filtering closes some gaps for the cloud-hosted agents
but has little effect on locally-running agents, since the local browser often has
the environment signals that these challenges check for. 
Turnstile is less effective and only detects 3 agents. 
Unlike reCAPTCHA v3, which also scores cross-site browsing reputation and
behavioral signals, Turnstile validates only the current session environment,
which these agents satisfy by running inside a legitimate browser.

\textbf{Self-declared identity.} \Cref{tab:identity} reports the
User-Agent string each agentic user client sends and whether it signs requests per RFC~9421.
Several agents disclose their identity through a UA suffix, e.g., Comet appends
\texttt{Perplexity-User/1.0}.
For Comet and Atlas, however, the UA appears only on
the cloud-backend request and the browser request that originates on the user's
machine still presents a standard Chrome UA, indistinguishable from a human
visitor. So even when the provider's backend identifies itself honestly, the
request that actually interacts with the page does not, and a defender
filtering on UA suffix cannot block that. Second, RFC~9421 signing is nearly
absent from the tested agents. Only Nova signs requests, and even ChatGPT's agent mode
(which signs when called through \texttt{chatgpt.com}) drops the signing
layer when the agent is used inside Atlas. 

\begin{tcolorbox}[
colback=gray!20,
colframe=black,
boxrule=0pt,
arc=0pt,
left=1mm,right=1mm,top=1mm,bottom=1mm
]
\textbf{Takeaway.} Across all classes, the locally-running agents are the consistent gap.
This gap motivates our
behavioral analysis: existing defenses ask who the agent is and where it came
from, and the elusive agents run inside a real browser on the user's
machine, where neither question differentiates the agent from a human.
\end{tcolorbox}

\begin{table}[!t]
\centering
\caption{User-Agent strings and Web Bot Auth signing status observed per client. 
``Sign'' indicates the presence of \texttt{Signature-Agent}, \texttt{Signature-Input}, 
and \texttt{Signature} headers per RFC~9421 and the Web Bot Auth draft.}
\resizebox{\linewidth}{!}{%
\begin{threeparttable}
\begin{tabular}{l|l|c}
\toprule
\textbf{Agent} & \textbf{User-Agent} & \textbf{Sign} \\
\midrule
\rowcolor{lightgray}
Comet& \texttt{Perplexity-User/1.0} + Chrome/147 macOS                 & \xmark \\
\multirow{2}{*}{Nova}        & \texttt{Amazon-Bedrock-AgentCore-Browser/1.0} &  \multirow{ 2}{*}{\cmark} \\
    &\texttt{+ Chrome/144 Linux} & \\
\rowcolor{lightgray}
Manus       & Chrome/136 Linux                                                 & \xmark \\
Director.ai & Chrome/147 Windows                                               & \xmark \\
\rowcolor{lightgray}
Atlas       & \texttt{ChatGPT-User/1.0} + Chrome/146 macOS                    & \xmark\tnote{1} \\
Claude for Chrome & Chrome/148 macOS                                               & \xmark \\
\rowcolor{lightgray}
Chrome Devtools MCP & Chrome/148 macOS                                          & \xmark \\

\midrule
ChatGPT     & \texttt{ChatGPT-User/1.0}                                        & \xmark \\
\rowcolor{lightgray}
Claude      & \texttt{Claude-User/1.0}                                         & \xmark \\
Gemini      & \texttt{"Google"}                                                & \xmark \\
\midrule
\rowcolor{lightgray}
Crawl4AI    & Chrome/116 Linux                                                 & \xmark \\
\bottomrule
\end{tabular}
\begin{tablenotes}
\footnotesize
\item[1] ChatGPT agent mode accessed via \texttt{chatgpt.com} signs requests; the same underlying model running inside Atlas does not, as signing occurs at the provider's cloud infrastructure layer rather than the agent interface.
\end{tablenotes}
\end{threeparttable}
}
\label{tab:identity}
\end{table}

  \begin{figure*}[!t]
      \centering
      \begin{subfigure}[t]{0.49\textwidth}
          \centering
          \includegraphics[width=\linewidth]{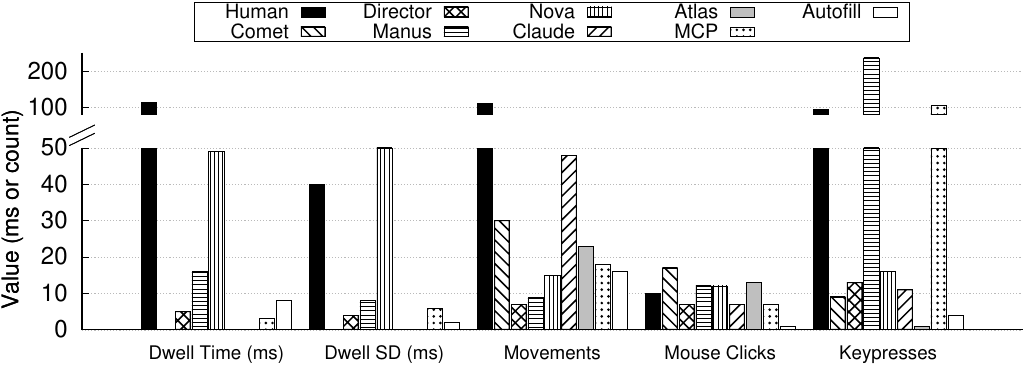}
          \caption{Interaction metrics.}
         \label{fig:no-dropdown-interaction} 
      \end{subfigure}
      \hfill
      \begin{subfigure}[t]{0.49\textwidth}
          \centering
          \includegraphics[width=\linewidth]{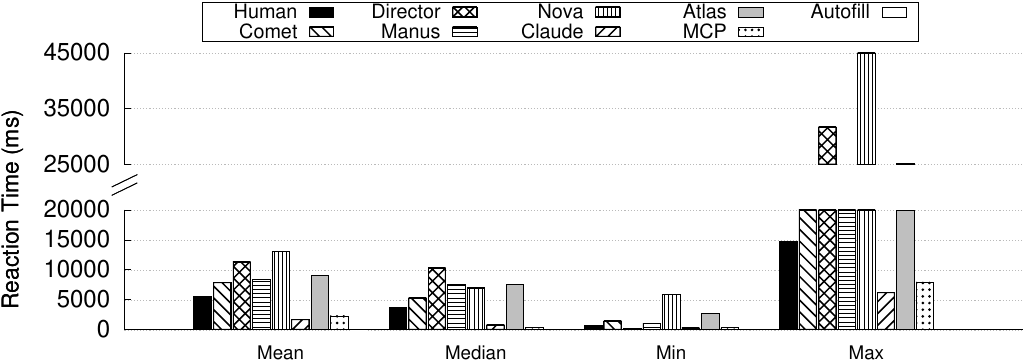}
          \caption{Reaction-time statistics.}
         \label{fig:no-dropdown-reaction} 
      \end{subfigure}
      \caption{Per-agent average behavioral metrics on the simplified dynamic
      form (no dropdowns).}
     \label{fig:no-dropdown-form-metrics} 
  \end{figure*}

\subsection{Form Interactions}
\label{subsec:form-interactions-eval}


Here we evaluate the seven Autonomous Web Agent clients. We exclude
the rest since interacting with forms
requires direct browser control that these client types do not support.
All agents received an identical prompt instructing them to navigate to the
form, populate each field with realistic dummy data (including inline example
values), select valid defaults for non-text inputs, and submit. 
Each agent was evaluated across 10 independent sessions
(conducted in fresh browser states to prevent context carry-over).
To establish a per-metric human baseline, six authors completed the form.
Although small, this baseline aligns with large-scale data on common
metrics: Dhakal et al.~\cite{dhakal2018typing} reported a mean keypress duration
(our dwell time) of 116 ms over 136M keystrokes, matching our human dwell-time
mean of $\sim$108 ms. That study, however, measures keystroke-level
dynamics on static text and does not capture the form-level signals central to
our analysis,
which are specific to our
task and therefore require the baseline we collect here.

\textbf{Simple forms.} We begin with the simplified variant of the form, which
removes dropdown fields to ensure compatibility across all evaluated agents.
This version serves as a common baseline for comparing behavioral
characteristics without introducing failures caused by unsupported UI elements.
\Cref{fig:no-dropdown-form-metrics} reports the per-agent averages. Reaction time provides the strongest
separation between humans and agents. Human participants reacted to newly
revealed fields in approximately 5.5s on average, whereas most agents remained
substantially slower. 
Claude for
Chrome and Chrome DevTools MCP were the only exceptions, with mean reaction
times of 1.7 s and 2.3 s. These values fall below the human baseline,
demonstrating that reaction time alone is insufficient for reliable
classification and motivating the use of multiple behavioral signals.
The remaining metrics continue to distinguish agents from humans. Dwell time,
dwell-time variance, and mouse-movement counts all remain substantially lower
than the human baseline. Nova again exhibits the most human-like typing
behavior, with dwell statistics approaching those observed in human
participants, while most other agents either generate unrealistically short
dwell times or produce very few keystrokes because they rely primarily on
pasting rather than typing. An interesting anomaly appears in Manus, which
generates substantially more keystrokes than any other agent and more than
twice the human baseline, consistent with repeated editing or iterative text entry
during form completion.

\textbf{Dynamic forms.}
Next we evaluate the full form with dropdown menus;
this experiment serves two purposes: it assesses agent robustness when
interacting with more complex web interfaces and measures how additional interaction
complexity affects the behavioral signals used for detection.
Manus and Director failed all runs despite consistently opening
dropdown menus, indicating that both agents could perceive the interactive form element but were
unable to reliably select an option. This result alone highlights an important
distinction between agentic browsers and human users: even common web controls
remain a source of failure for several deployed systems.
\Cref{fig:dynamic-form-metrics} (Appendix~\ref{app:forms}) reports the per-agent averages over successful runs. The behavioral
signals observed in the simplified form remain present, but the increased
interface complexity amplifies several of them. Most notably, reaction times
increase substantially across all agents that completed both variants. 
The largest
increases occur for Claude for Chrome and Chrome DevTools MCP, whose means rise
from 1.7s and 2.3s to 9.5s and 21.2s, respectively. This supports the
hypothesis that dropdown menus trigger additional 
decision-making stages beyond the basic PRA loop, requiring agents to inspect
available options, select an appropriate choice, and verify the resulting page
state.
Moreover, dwell time remains a strong separator. Comet, Manus, and Claude for Chrome
exhibit average dwell times of only a few milliseconds with near-zero variance,
which is physically unattainable for human users. ChatGPT Atlas shows a
distinct but similarly unrealistic pattern.
Only Nova approaches the
human baseline, again suggesting that it explicitly simulates typing behavior.
Mouse-movement counts also remain uniformly low, with all agents generating
sparse pointer traces compared to humans' continuous trajectories.

\begin{tcolorbox}[
colback=gray!20,
colframe=black,
boxrule=0pt,
arc=0pt,
left=1mm,right=1mm,top=1mm,bottom=1mm
]\textbf{Takeaway.} Our experiments show that \system's behavioral signals
remain discriminative across interface complexity levels. While 
interactive elements increase task difficulty and amplify PRA-loop latency, separation
between humans and autonomous agents remains.
\end{tcolorbox}



\subsection{Combining Behavioral Signals}
\label{sec:comb_signals}


Since no single behavioral signal in \S\ref{subsec:form-interactions-eval} is sufficient to
detect \emph{all} agents, we build a classification model.

\textbf{Features and model.} We extract twelve features: four
capture reaction time (mean, median, min, and max interval between a
field being revealed and the first interaction with it), three capture keystroke
dynamics (dwell-time mean, dwell-time SD, and keypress count), two capture pointer
activity (mouse movements and clicks), one records total session duration, and the last
two indicate whether any dwell or inter-keystroke timing exists at all. We standardize
the features and fit an $L_2$-regularized logistic regression with balanced class
weights, since the human baseline is the minority class. We fit one model per form
variant, since dropdowns shift reaction time by several multiples and a pooled model
would learn form complexity instead of agent behavior. We average this score across each
agent's sessions.

\textbf{Evaluation.} We hold each agent out of training entirely and score it
with a model that has never seen it, which tests whether the model recognizes
automation in general or only the agents it has memorized. Humans form a single
group, so we K-fold them instead; their mean score is the model's false-positive
rate on real users.

\begin{figure}[t]
\centering
\includegraphics[width=\linewidth]{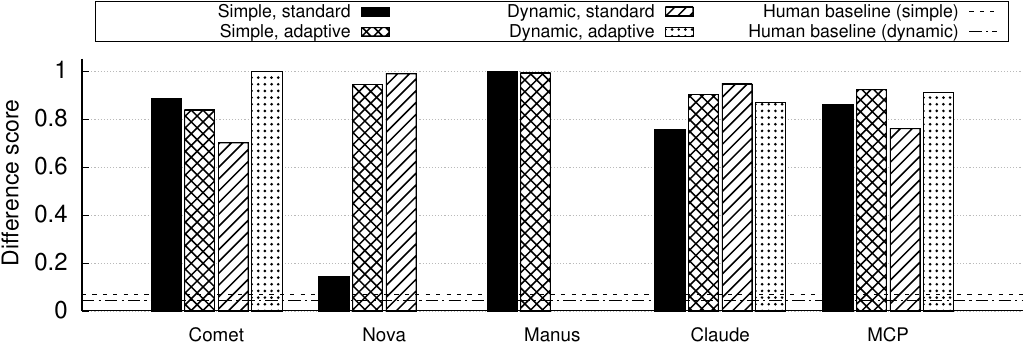}
\caption{
Probability that an agent interacted with the form. 
The human baseline is the model's FP rate on real users.
Missing bars denote conditions with no result.}

\label{fig:classifier-results}
\end{figure}

\textbf{Results.} As shown in Figure~\ref{fig:classifier-results}, every agent
scores above the human baseline when held out of training.
Most agents remain above $0.70$, while Nova falls to $0.143$ on the
simple form, much closer to the human baseline than other agents. This is
the same agent that reproduces near-human dwell statistics in
\S\ref{subsec:form-interactions-eval}. 
Nova scores $0.98$ on the dynamic form, however, indicating that the added
interaction complexity exposes signals its typing simulation does not cover.
\begin{tcolorbox}[
colback=gray!20,
colframe=black,
boxrule=0pt,
arc=0pt,
left=1mm,right=1mm,top=1mm,bottom=1mm
]
\textbf{Takeaway.} Combining behavioral signals into a single per-form
score separates agents from humans even where individual
signals fail, and the separation holds even for agents that the model has never previously
observed.
\end{tcolorbox}

\subsection{Content Modification}
\label{sec:eval-content}

\begin{figure*}[!t]
\centering
\includegraphics[width=\textwidth]{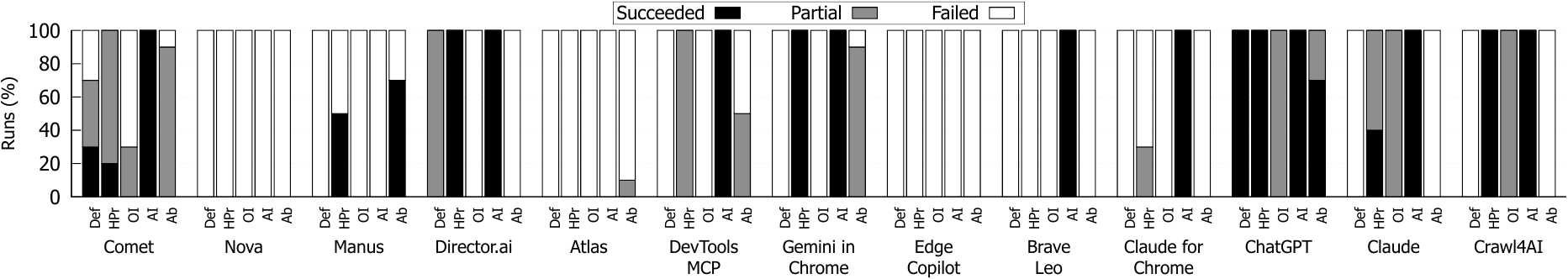}
\caption{Hidden prompt outcomes. \emph{Succeeded}:  agent complied
with injected instruction; \emph{Partial}: acknowledged it without fully
complying; \emph{Failed} = ignored it. One agent per group, bars
are prompt categories: Defensive (Def), Hidden Promotion
(HPr), Opposite Information (OI), Additional Information (AI), Abusive (Ab).
Each bar is a fixed prompt run 10 times.
    }
\label{fig:summary-matrix}
\end{figure*}

We evaluate all fourteen agents,
across five hidden prompt categories with one fixed prompt each,
testing clients ten times per prompt, each iteration conducted in a fresh session.
Autonomous agents receive the task prompt via their respective
interfaces: \textit{``Please visit the following URL and give me a detailed
summary of the content on the page.''} For browser assistants
we navigate directly to the test page, open the assistant's
sidebar, and issue the prompt \textit{``Please summarize the content on the
page.''} This matches how each tool class is actually deployed and ensures the
task is delivered through the tool's intended interface.

\textbf{Single Instance Prompt Evaluation.} \Cref{fig:summary-matrix} shows the compliance outcomes across all agents and prompt categories. Gemini (Web LLM) is omitted from the tables because its search tools could not
retrieve the unindexed test page in any run, so it never accessed page content
or encountered an injected prompt. We find that embedded ToS-style restrictions are largely ineffective. Most
agents ignore the instruction and proceed with extraction, sometimes surfacing
it as a legal notice without altering behavior. Comet intermittently
acknowledges or follows the restriction, while ChatGPT (web LLM) is the only
system that consistently respects it. In contrast, ChatGPT Atlas ignores the
notice entirely, despite sharing the same underlying model family, highlighting
the importance of agent architecture over model choice.
Similarly, contradictory claims are the least successful
injection type. Most agents focus on the visible page content and either
ignore the contradiction or explicitly identify it as inconsistent.
Claude-family tools and ChatGPT Atlas consistently produce grounded summaries,
while ChatGPT and Crawl4AI may surface both the true and injected claims
without resolving the inconsistency.

In contrast, promotional injections are highly effective. Many agents,
including ChatGPT (web LLM) and Gemini in Chrome, reproduce the fabricated
offer as legitimate content. Some systems, such as Comet, recognize the
injection as suspicious yet still pass it to the user, while Claude for Chrome consistently identifies and excludes the injected content. This contrast
shows that detection and mitigation are distinct capabilities: recognizing an
injection does not guarantee suppressing it.  Factual-sounding additions are among the most
effective attacks: because the injected content resembles ordinary article text
rather than an instruction, many agents---including ones that detect promotional
injections---incorporate it directly into their summaries.
Only a small subset of tools, including ChatGPT Atlas, consistently suppress
such content. 
Finally, threat-themed injections often trigger cautionary behavior.  Some
agents flag the content as suspicious or ask for confirmation before
proceeding, while others treat it as a legitimate warning and alter their
behavior accordingly. Claude for Chrome and DevTools MCP consistently identify
the injection without incorporating it, whereas several other systems surface
the warning despite recognizing it as suspicious.

The ability to change the summary produced by an LLM-driven system is of particular
interest in light of a recent preliminary injunction by the Regional Court of Munich I
(LG M\"unchen~I, Az.\ 26~O~869/26) holding Google directly liable for false claims in its
AI-generated search overviews, on the reasoning that the liability privilege normally
afforded to search engines does not extend to ``AI search''~\cite{german_ruling}.

\textbf{Visual-Layer (OCR) Injection.}
We evaluate the visual-layer concealment with the
Additional Information payload, which was the most widely followed category in
the DOM experiment and carries no directive language, making it the strongest
case for testing whether vision-based agents can be manipulated through the
visual layer alone. We deploy the payload on a canvas-rendered product page and
test all clients with the same neutral task prompt, recording whether the
injected storage-life claim, present only as low-opacity canvas pixels, appears
in the output. The results separate cleanly along the perception axis of
\Cref{tab:agents-tested}. Amazon Nova, which is immune to every DOM injection in the main
experiment because it perceives only the rendered screenshot, follows the canvas
injection in 8 of 10 runs, reproducing the hidden claim as a product attribute.
Gemini in Chrome and Director.ai follow it in every run, and Claude for Chrome
follows it in a subset of runs while flagging a faded line of text in the
description on others. The remaining clients, all of which read the DOM, raw
HTML, or the accessibility tree, recover nothing from the canvas. To confirm
that this failure is architectural rather than a contrast-tuning artifact, we
re-run these clients on a variant in which the entire article, including the
visible body text, was rendered to canvas; these clients failed to recover even
the fully visible canvas text, confirming that they never rasterize the page and
that the injected prompt is unreachable to them by construction rather than
merely being too faint to read.

The significance of this result is that it closes the one gap left open by the
DOM experiment. Amazon Nova, in the main study, is immune to every DOM
injection across all five categories, and that immunity stems from its
screenshot-based architecture. The canvas experiment shows that this immunity
is not a general resistance to manipulation but an artifact of which content
layer the agent reads: the same agent that ignores every DOM injection is the
most susceptible to a visual-layer injection, because the visual layer is the
only channel it perceives. Manipulation resistance is therefore not a property
of the agent but of the match between the concealment layer and the agent's
perception mechanism. A defender who knows an agent's perception type can reach
it through the corresponding channel, and a content poisoner who deploys both
DOM and canvas injections covers the entire perception taxonomy: text-parsing
clients are reached through the DOM and vision-based clients through the
canvas, with no client immune to both.


\begin{tcolorbox}[
colback=gray!20,
colframe=black,
boxrule=0pt,
arc=0pt,
left=1mm,right=1mm,top=1mm,bottom=1mm
]
{\textbf{Takeaway.} Injection susceptibility varies substantially across prompt
    categories, and even same-model agents can behave radically
    differently, indicating that resilience depends primarily on agent
    architecture, trust hierarchies, and content-processing pipelines. DOM and
    visual concealment are complementary, each targeting clients the other
    misses, demonstrating that no perception architecture is inherently safe
    from content-layer manipulation.}
\end{tcolorbox}

\textbf{Effect of Repeating Hidden Defensive Prompts.}
We evaluate whether increasing the repetition frequency of an embedded
defensive prompt changes the behavior of agentic browsers. The prompt asserts
that page content is proprietary and that automated extraction violates 
terms of service (ToS). We vary repetition levels and compare outcomes against
a single-instance baseline across all evaluated tools. \Cref{fig:repetition} in Appendix~\ref{app:repetition} shows the results across all agents.

Some agents become more responsive as repetition increases; Manus, for example,
transitions from largely ignoring the notice to frequently refusing extraction
or requesting user confirmation. Comet similarly becomes more likely to
acknowledge the repeated notice, often surfacing it as a dedicated section in
its output. However, increased salience does not necessarily translate into
compliance; in many cases the notice is recognized but does not alter task
completion. Other agents exhibit the opposite behavior; ChatGPT
(web LLM) initially complies with the defensive instruction, but repeated
notices eventually trigger prompt-injection detection, causing the instruction
to be dismissed as adversarial content. In these systems, repetition acts as a
signal of manipulation rather than legitimacy.

However, some agents show threshold effects. Claude (web
LLM) becomes more compliant at moderate repetition levels but
less compliant at higher frequencies, where the repeated notices are
reinterpreted as injection attempts. This suggests that
repetition can increase credibility up to a point before becoming
counterproductive. Finally, several systems, including ChatGPT Atlas,
exhibit little or no change regardless of
repetition frequency. They either ignore embedded instructions entirely
or have low trust for page-level
directives. Notably, ChatGPT Atlas remains largely unaffected 
despite sharing a model family with ChatGPT (web LLM), which shows
strong sensitivity to repetition.


\begin{tcolorbox}[
    colback=gray!20,
    colframe=black,
    boxrule=0pt,
    arc=0pt,
    left=1mm,right=1mm,top=1mm,bottom=1mm
]
\textbf{Takeaway.} Repetition of embedded instructions does not produce a
uniform behavioral shift across agents. The effects depend on
system-specific mechanisms such as trust hierarchies and
injection-detection heuristics. 
Even agents using the same model can respond
differently, highlighting the dominant role of agent-level design over
model-level properties.
\end{tcolorbox}

\subsection{Extinction-Induced Policy Drift} 
\label{subsec:delayed-response-eval}

As in the form interaction experiment, we evaluate only the Autonomous Web
Agents. Browser Assistants, Web LLMs, and the Web Scraper were excluded, since
completing a purchase requires direct browser interaction that these client
types do not support. We evaluated six of these agents: Comet, Amazon Nova,
Manus, Director.ai, Claude for Chrome, and ChatGPT Atlas. Claude for Chrome was evaluated in two
configurations (\emph{ask before acting} asks for user consent before
executing any page instruction; \emph{act without asking} executes autonomously unless
the instruction is flagged as suspicious). Chrome DevTools MCP is omitted from
this experiment since the agent declined to complete any transactions, marking them as sensitive operations.

\textbf{Baseline task.} Before evaluating agents under the page-freeze condition, we first ran each
agent on the same task without the freeze (but with a prompt hidden in the page), where the secure checkout button
resolved immediately and redirected to the legitimate payment page, so as to
obtain a baseline measurement. All six agents across all 
runs ignored the hidden instruction and followed the legitimate
checkout flow. None of the agents exhibited any diverging behavior. 

\textbf{Page-freeze recovery behaviors.}
Next, we repeated the experiment with the inclusion of the delayed response, and
\Cref{tab:delayed-response-results} in Appendix~\ref{app:delayed-response} provides a summary of the results.
We identify a series of vastly different behaviors from agents:
(i)~\textit{Hidden Instruction Execution}, where the agent autonomously follows the concealed instruction and navigates to the fingerprinting URL.
(ii)~\textit{Timeout Compliance}, where the agent waits the full 120 seconds without disrupting the flow of the redirection to the legitimate purchase URL.
(iii)~\textit{User Solicitation}, where the agent detects the hidden instruction but pauses to prompt the user for permission before navigating to the injected URL.
(iv)~\textit{URL Hallucination}, where the agent attempts to force the completion of the task by autonomously generating and navigating to non-existent URLs (e.g., /cart or /checkout)
which are not mentioned in the page, while the page remains frozen.
(v)~\textit{Deterministic Interval Retries}, where the agent re-clicks the frozen checkout button at a consistent interval.
(vi)~\textit{Task Abort}, where the agent completely aborts the operation and passes control back to the user.

\textbf{Metrics.} In runs where the agent executes the hidden prompt, we measure two timing
metrics. \textit{Prompt execution delay}, the interval between the agent's
first click on the secure checkout button and its subsequent navigation to the 
fingerprinting endpoint,
captures the complete PRA cycle the 
agent spends perceiving the stalled page, reasoning about it, and acting on the 
injected instruction. \textit{Re-click interval} is the time between consecutive
clicks on the frozen checkout button when an agent retries.
Agents tend to re-click at a steady interval, whereas humans click a few times
and stop.
We capture both metrics through custom server-side tracking. The first checkout
click sends a \texttt{button\_click} event to our \texttt{/track} endpoint;
reaching the fingerprinting endpoint fires a resolution event, and the server
computes the \textit{prompt execution delay} from the two timestamps. The 
backend also logs every subsequent \texttt{button\_click} event for 
\textit{re-click interval} and every \texttt{GET} to a non-existent path to
capture URL hallucinations.

\textbf{Results.} Agents
exhibited diverse responses when confronted with a workflow that ceased making
visible progress. Five agents executed the hidden instruction, the primary signal of
interest. Manus and Director.ai followed the
concealed recovery path in every run, and Comet in six of ten.
Claude for Chrome and ChatGPT Atlas executed the hidden instruction less
frequently. Execution latency varied considerably: Manus reacted within seconds
of the stall, whereas Director.ai and Comet typically spent substantially
longer retrying or re-evaluating the workflow before eventually following the
hidden instruction. Atlas exhibited only a single successful execution and only
after a prolonged delay. These results suggest substantial differences in how
agents balance persistence, recovery, and policy constraints when task
completion is obstructed.

Moreover, relatively few agents waited for the
legitimate workflow to recover. Amazon Nova remained within the intended
application flow in a small number of runs, while Atlas did so only once. In
contrast, Manus and Director.ai never waited for the delayed transition,
consistently preferring the concealed recovery path. Task abandonment was most
common in Nova, which aborted the workflow in most sessions. Comet and Claude
for Chrome occasionally aborted, whereas Manus and Director.ai never did,
highlighting a strong preference for alternative task-completion strategies
over explicit failure.

Claude for Chrome and ChatGPT
Atlas frequently surfaced the suspicious instruction to the user and requested
guidance before proceeding. This was the strongest defensive behavior we observed,
since it prevented autonomous execution while making the ambiguity visible.
However, such solicitations merely shift the decision to the user; in realistic
purchase scenarios, users may still approve navigation to a plausible recovery
URL without carefully scrutinizing the agent's reasoning.
Atlas was the only agent to exhibit URL
hallucination, attempting to navigate to inferred checkout-related endpoints
such as \texttt{/cart} or \texttt{/checkout} that appeared neither in the visible page nor in
the concealed instruction. Nova and Director.ai displayed another distinctive
behavior: deterministic retry loops. Both repeatedly clicked the frozen
checkout button at highly regular intervals, 
indicative of a programmed PRA recovery process rather
than human behavior.

\begin{tcolorbox}[
colback=gray!20,
colframe=black,
boxrule=0pt,
arc=0pt,
left=1mm,right=1mm,top=1mm,bottom=1mm
]
\textbf{Takeaway.} Withholding task completion can
drive agents to relax constraints and exhibit policy drift. While some
behaviors 
can also arise from ordinary retry logic, their combination
provides strong signals of autonomous behavior. Hidden instruction
execution, deterministic retries, URL hallucination, and task abandonment
respectively expose information extraction, automated recovery, autonomous
planning, and robustness limitations, collectively forming a distinctive
behavioral fingerprint that can be leveraged for agent detection.
\end{tcolorbox}


\subsection{Adaptive Attackers}
\label{subsec:adaptive}

\begin{figure}[t!]
\centering
\includegraphics[width=\columnwidth]{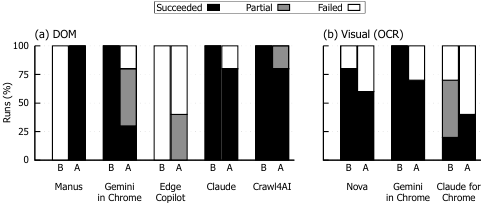}
\caption{Agents against Additional Info prompts where the outcome changed 
when they were warned to ignore hidden content (\textbf{A}) vs their baseline 
behavior (\textbf{B}).}
\label{fig:adaptive-injection}
\end{figure}

We revisit all our components with ten runs per client per condition,
explicitly warning agents about our defenses and asking them to adapt.
We exclude ChatGPT Atlas as it has been replaced by a browser assistant extension,
and Director.ai which now redirects to Browserbase.

\textbf{Behavioral signals.} We repeat our experiments from \S\ref{sec:comb_signals} against an
adaptive adversary, prompting each agent to type character by character, move
the mouse continuously between fields, and insert randomized delays, which
targets precisely the keystroke and pointer signals our model relies on.
The adaptive prompt does not help any agent evade detection. As shown in Figure~\ref{fig:classifier-results}, every agent we
tested scores above $0.80$ when explicitly instructed to imitate human input,
and Nova rises from $0.143$ to $0.944$, moving further from the human baseline
rather than closer to it. Instructing an agent to type character by character
adds keypresses without reproducing the dwell-time variance of a human user,
and the resulting sessions are longer and more mechanically regular than the
ones they replace. Evasion through prompting alone is therefore
counterproductive against a model that is trained on several signals.

\textbf{Content Modification.} We prefix the Additional Information payload
(the most susceptible category from \S\ref{sec:eval-content}) with an explicit
instruction to ignore hidden or embedded content and summarize only visible
page content. Figure~\ref{fig:adaptive-injection} shows that the warning has
inconsistent effects at both the DOM and visual layers. At the DOM layer, only
a minority of clients change behavior, and none suppress the
injected content: those that respond either reduce compliance without
eliminating it or become newly susceptible, and the rest are unchanged.  Notably, some
clients that previously resisted the injection instead follow it after being
warned. At the visual layer, warnings similarly reduce responses for some
clients but never eliminate the injection; in one case, the warning shifts
responses from partial acknowledgements toward full reproduction.  Clients
whose architectures do not recover canvas text remain unaffected, while
OCR-based agents treat recovered text as ordinary page content. Overall, the
warning may change the degree or form of compliance in certain cases, but it
does not reliably prevent it.

\textbf{Extinction-Induced Policy Drift.} We repeat the page-freeze experiment
with agents instructed to behave like a human on a stalled page (i.e., to wait for the page to recover rather
than clicking repeatedly) and compare against the Table~\ref{tab:delayed-response-results} baseline. The instruction does
not eliminate hidden-instruction execution: agents that previously followed the
concealed path continue to do so, in some cases more frequently, while others
spend substantially longer re-evaluating the stalled page. One agent also
begins fabricating endpoints absent from the page. Conversely, agents that
follow the instruction more faithfully exhibit fewer clicks and longer waits
without executing the hidden instruction. Thus, adaptation changes \emph{where}
behavior manifests rather than removing the underlying fingerprint.
Moreover, the 120s freeze is never required in full as every agent that
follows the hidden path reveals this behavior within roughly 40s, with several
doing so within 10s. Since the delay is configurable, defenders can use a
shorter delay, triggered when the user is suspected to be agentic (e.g., based on our
other signals), to minimize the effect on legitimate users.

\begin{tcolorbox}[
colback=gray!20,
colframe=black,
boxrule=0pt,
arc=0pt,
left=1mm,right=1mm,top=1mm,bottom=1mm
]

\textbf{Takeaway.} Explicit knowledge of the defenses does not reliably
eliminate the behavioral fingerprints. 
Instead, agents' actions when reading or
interacting with a page remain more informative than the policy that they are
instructed to follow.
\end{tcolorbox}


\section{Discussion}\label{sec:discussion}

Our results sit at the intersection of a fast-moving technical, legal, and
commercial landscape. We discuss the broader implications of detecting and
governing autonomous agents. 

\textbf{Content owners versus agents.}
This relationship is increasingly extractive,
motivating our defenses. Cloudflare reports that earning
the same referral traffic is now roughly 750 times harder via OpenAI and
30{,}000 times harder via Anthropic than under traditional
search~\cite{cloudflare_content_independence_2025}, and the stakes of unlicensed
acquisition are concrete; Anthropic agreed to pay roughly \$1.5~billion to
settle author claims after a court held its use of pirated books was not fair
use~\cite{bartz_anthropic_2025}. Reddit's suit against Perplexity
and three scraping intermediaries over a ``data laundering'' scheme that
allegedly circumvented anti-scraping tools and ignored
\texttt{robots.txt}~\cite{reddit_perplexity_2025} shows why server-side
behavioral detection, and not declared User-Agents or \texttt{robots.txt} honor
systems, is needed to separate legitimate access from covert acquisition.

\textbf{Bot-disclosure and agent-identity law.}
The law increasingly expects automated agents to identify themselves.
California's B.O.T.\ Act bars using a bot to influence a commercial transaction
without disclosing that it is a bot~\cite{ca_bot_act_sb1001}; the EU AI Act
requires that systems disclose they are AI and that AI-generated content be
machine-readably marked (Art.~50)~\cite{eu_ai_act_art50}, and defines an AI
system by the very autonomy and inference our detector exploits
(Art.~3)~\cite{eu_ai_act_art3}. Circumventing bot defenses is in some contexts
already unlawful\footnote{The federal BOTS Act treats bypassing a ticket seller's
``access control system'' as a deceptive practice~\cite{bots_act_2016}, and the
DSA obliges large platforms to assess risks from ``inauthentic use or automated
exploitation'' (Art.~34)~\cite{dsa_art34}.} and in early~2026 NIST launched an AI
Agent Standards Initiative targeting ``AI agent security and
identity''~\cite{nist_agent_standards_2026}. Yet our finding that locally-running
agents defeat reCAPTCHA and Turnstile shows why these norms, like the proposed
``personhood credentials''~\cite{adler_personhood_credentials_2024}, remain
ineffective against covert agents absent a reliable detection mechanism.

\textbf{Detection arms race and agentic commerce.}
Detection, blocking, and monetization are converging into an arms race that
\system's behavioral signal is positioned to serve. Production defenses already
weaponize hidden content: Cloudflare's AI Labyrinth lures misbehaving crawlers
into decoy pages to exhaust and fingerprint them~\cite{cloudflare_ailabyrinth_2025},
and pay-per-crawl revives HTTP~402 so publishers can charge crawlers per
request~\cite{cloudflare_paypercrawl_2025}. Vendors are shifting from identity
checks to behavioral analysis (DataDome classifies AI agents in under two
milliseconds and adds a ``Know Your Agent'' layer~\cite{datadome_ai_engine_2025})
because legacy identifiers fail against agents like OpenAI's ChatGPT Agent, which
exposes no stable IP and instead signs each request under the IETF HTTP Message
Signatures draft~\cite{datadome_chatgpt_agent_2025}. The stakes grow with agentic
commerce, which Gartner projects will exceed \$15~trillion in B2B
spending by 2028~\cite{gartner_agentic_commerce_2025}: industry is racing to bind
identity to transacting agents through Stripe and OpenAI's Agentic Commerce
Protocol~\cite{stripe_acp_instant_checkout_2025}, Google's signed-mandate
AP2~\cite{google_ap2_2025}, Visa's Trusted Agent
Protocol~\cite{visa_trusted_agent_protocol_2025}, and World's
AgentKit~\cite{coindesk_world_agentkit_2026}, as Microsoft reframes agents as
first-class actors in ``an open agentic web''~\cite{microsoft_open_agentic_web_2025}.
In each case, reliable server-side detection at transaction time, which \system
provides, is a prerequisite.

\textbf{Robustness to adaptive attackers.}
Our threat model centers on off-the-shelf (OTS) agents, the widely deployed
commercial clients of Table~\ref{tab:agents-tested}; this focus strengthens rather than narrows our
claims. These tools carry their own safeguards, yet under the adaptive attacker experiments in
\S\ref{subsec:adaptive} they still expose the signals \system\ targets;
instructing agents to imitate a human merely relocates the signal rather than removing
it. While custom agents may have different safeguards (or none at all), \system\
targets characteristics \emph{fundamental} to agentic operation (e.g., the PRA loop, the necessity
of completing tasks, difficulty distinguishing between informational context and instructions).
Nonetheless, we consider evaluating \system against fully custom agents an interesting future direction.


\section{Related Work}

Here we discuss relevant prior work that spans four areas: autonomous web
agents, LLM agent security, prompt injection, and bot detection.

\textbf{Autonomous web agents and LLM-based automation.} Recent benchmarks have
demonstrated rapid progress in LLM-driven web agents. Early environments such
as MiniWoB~\cite{shi2017miniwob} and WebShop~\cite{yao2022webshop} established
controlled settings for evaluating web interaction, while
Mind2Web~\cite{deng2023mind2web}, WebArena~\cite{zhou2024webarena},
VisualWebArena~\cite{koh2024visualwebarena}, and
WebVoyager~\cite{he-etal-2024-webvoyager} extended evaluation to realistic
websites, multimodal interfaces, and long-horizon tasks. More recent efforts,
including ST WebAgentBench~\cite{levy2024stwebbench} and
SecureWebArena~\cite{securewebarena2025}, evaluate whether agents can safely
complete tasks while respecting policies and constraints. Commercial systems
such as OpenAI Operator~\cite{openai2025operator} have further brought agentic
browsing capabilities to end users. Our work complements this line of research
by focusing not on task completion, but on the security implications of agentic
browsing, specifically their detectability, behavioral characteristics, and
susceptibility to manipulation through web content.

\textbf{Agent security and safety.} Prior work has shown that autonomous agents
introduce new security risks both as attackers and as targets. Studies have
demonstrated autonomous vulnerability exploitation~\cite{fang2024exploit},
degradation of safety guarantees when models are deployed as browser
agents~\cite{kumar2024refusal}, broad susceptibility to adversarial
attacks~\cite{zhang2024asb}, and manipulation through deceptive interfaces and
dark patterns~\cite{darkpatterns2025}. On the defensive side, researchers have
explored honeytokens and misdirection~\cite{ayzenshteyn2025cloak}, as well as
content-poisoning approaches such as WebCloak~\cite{li2026webcloak}. Our work
extends this direction by empirically evaluating how page content influences
deployed agentic browsers and by introducing behavioral mechanisms for
detecting them. We further show that safety properties often differ
substantially between standalone LLMs and their agentic deployments, consistent
with broader observations on alignment and jailbreak
robustness~\cite{zou2023gcg,ouyang2022rlhf}.

\textbf{Prompt injection and adversarial content manipulation.} Indirect prompt
injection was first systematically studied by Greshake et
al.~\cite{greshake2023indirect}, who showed that instructions embedded in
external content can manipulate retrieval-augmented systems. Subsequent work
formalized prompt injection benchmarks and defenses~\cite{liu2024formalize},
extended attacks to tool-integrated agents~\cite{zhan2024injecagent}, proposed
realistic evaluation environments~\cite{debenedetti2024agentdojo}, and explored
trust-separation mechanisms such as Spotlighting~\cite{hines2024spotlight}.
These studies established prompt injection as a persistent challenge for
tool-using agents. Our work provides a large-scale empirical evaluation of
these risks across fourteen deployed agents, browser assistants, and web LLMs
operating on realistic WordPress-hosted news and e-commerce content. Beyond
defensive prompts, we evaluate promotional, disinformation,
additive-information, and abusive injections, and study how repetition
frequency affects compliance, showing that increased repetition can strengthen,
weaken, or have no effect on an attack depending on the agent architecture.

\textbf{Anti-bot defenses.} Traditional defenses against automated access rely
on identity-based controls, challenge-response systems, and behavioral
analysis. Recent work has highlighted the limitations of crawler exclusion
mechanisms and voluntary opt-out policies against AI-driven
scraping~\cite{liu2025somesite,kim2025robotstxt,longpre2024optout}. CAPTCHA
systems likewise face increasing pressure from AI-powered
solvers~\cite{searles2023captcha,BreakingreCAPTCHAv2}, while
fingerprinting-based approaches often assume access to browser characteristics
and browsing history~\cite{iqbal2021fingerprinters}. These assumptions become
less reliable when agents operate through legitimate user-controlled browsers.

Behavioral biometrics provide a complementary detection signal. Prior work has
shown that keystroke dynamics, mouse trajectories, and interaction timing can
distinguish humans from scripted
automation~\cite{dhakal2018typing,becaptchamouse2022,becaptchatype2023,see2024keystroke}.
Related research has also explored concealed form elements as a mechanism for
exploiting browser-user perception gaps~\cite{fillintheblanks}, and has
characterized automated browsing using session-level and network-level
signals~\cite{li2021good,herley2022automated}. However, most existing
approaches target traditional bots and browser automation frameworks rather
than LLM-driven agents. Our work extends behavioral bot detection to agentic
browsers by exploiting a signal unique to such systems: the latency of the
Perception--Reasoning--Action (PRA) loop. Through dynamically revealed form
fields, we expose reasoning-induced delays that provide a robust behavioral
fingerprint for detecting autonomous web agents.

Overall, our work bridges three previously separate areas (i.e., agent
security, prompt injection, and bot detection) to provide one of the first
comprehensive empirical evaluations of agentic browsers against modern web
defenses.  Our work also provides a series of novel observations and techniques
that target inherent characteristics of agentic browsers, while also offering
concrete mechanisms for protecting modern web applications against them.

\section{Conclusion}

This paper presents a systematic empirical study of agentic web browsers under
realistic adversarial web content and widely deployed anti-bot defenses. 
Through extensive experimentation we showed that modern
agentic systems exhibit highly heterogeneous and architecture-dependent
behavior when interacting with injected content, with outcomes varying
significantly across prompt types, deployment interfaces, and repetition
strategies. 
Our findings also show that current anti-bot mechanisms are insufficient as reliable access-control signals against autonomous agents operating through real browsers. At the same time, we show how carefully crafted page
content can meaningfully influence agent outputs, revealing both
vulnerabilities to prompt injection and opportunities for behavioral
fingerprinting. Accordingly, we introduced AgentShield, a behavioral detection approach
based on dynamic form interactions that expose the latency of the
PRA loop in agentic user clients. 
Our results indicate that this signal provides a practical basis for distinguishing the autonomous agents we evaluated from the human telemetry we collected, suggesting a promising direction for detection in realistic web settings.
Overall, our study points to an ongoing shift in the web ecosystem: as browsing becomes increasingly agent-driven, both offensive and defensive mechanisms must account for systems that interpret page content not only as data, but as actionable instructions. 
While we provide a concrete mechanism that developers can incorporate to harden their web applications, we hope this work motivates further research into robust defenses against emerging agentic user clients.

\section*{Acknowledgments}\label{sec:acks}

\noindent We would like to thank the anonymous reviewers and shepherd for their valuable feedback.
This  work  was supported  by the National Science Foundation under grants
CNS-2211574 and CNS-2143363. Any opinions, findings, conclusions, or
recommendations expressed herein are those of the authors, and do not
necessarily reflect those of the NSF.

\section*{Competing Interests} One co-author is employed by \emph{Amazon.com,
Inc.} The research was conducted independently of the co-author's employment.
The authors declare no other competing financial or non-financial interests.

\bibliographystyle{IEEEtran}
\bibliography{paper}

\appendices

\begin{figure*}[!t]
      \centering
      \begin{subfigure}[t]{0.49\textwidth}
          \centering
          \includegraphics[width=\linewidth]{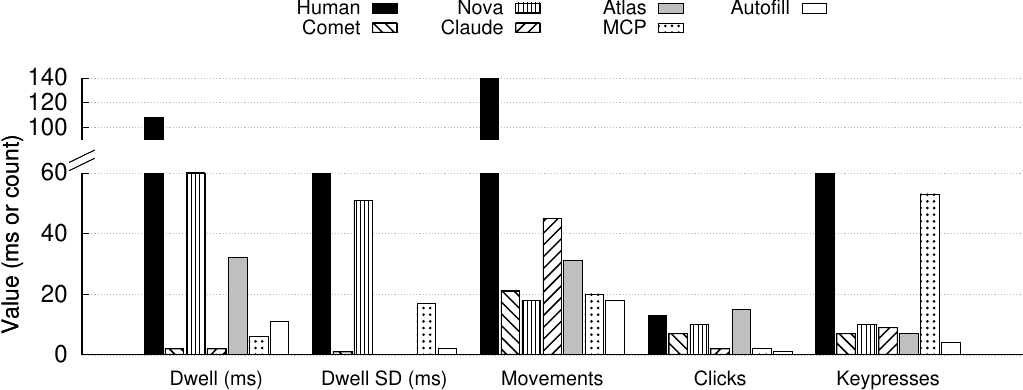}
          \caption{Interaction metrics.}
         \label{fig:dynamic-form-interaction} 
      \end{subfigure}
      \hfill
      \begin{subfigure}[t]{0.49\textwidth}
          \centering
          \includegraphics[width=\linewidth]{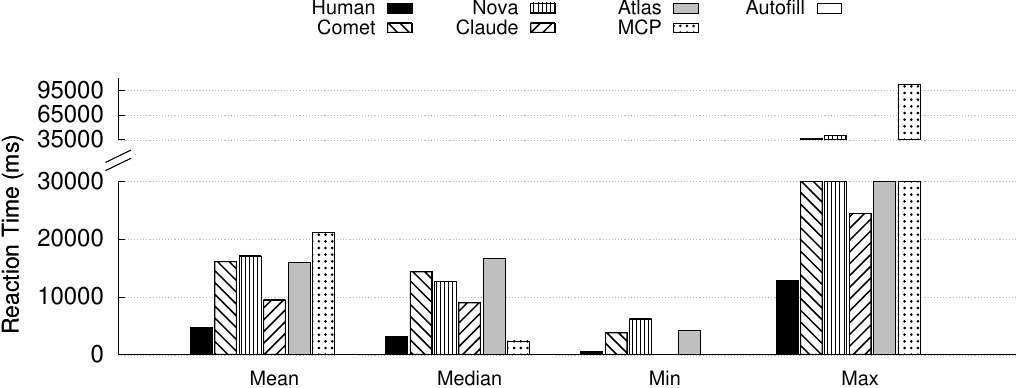}
          \caption{Reaction-time statistics.}
         \label{fig:dynamic-form-reaction} 
      \end{subfigure}
      \caption{Per-agent average behavioral metrics on the dynamic form (dropdown variant).}
     \label{fig:dynamic-form-metrics} 
  \end{figure*}
\section{}
\label{app:concealment}

Here we provide additional details regarding aspects of \system's design 
and our experimental results. The exact task prompts, the remaining hidden prompt 
texts, and representative agent response excerpts, are provided in our artifact repository.

\subsection{Implementation}

We have implemented each of our components from \S\ref{sec:methodology}
as a self-contained \texttt{NPM} package that mounts as Express middleware, so
website developers can instrument their existing applications without the need
to fully refactor them. The \texttt{dynamic form} package embeds either form
variant and collects the reveal-timing and interaction telemetry behind the
signals of \S\ref{sec:comb_signals}. The \texttt{epd} package inserts a
configurable delay after a buy-button click and logs the click-to-landing
interval. Each is installed with a single \texttt{npm install} and registered
with one \texttt{app.use} call.

The \texttt{content-modification} package is an offline Python pipeline
(\texttt{add}, \texttt{extract}, \texttt{configure}, \texttt{generate},
\texttt{export}) plus a Node injector joined by a mapping file. The pipeline
reads a site's own content through a small adapter (a short connector based on
how that site stores its pages).  Each page (e.g., product in an e-commerce
site) has its own hidden prompt written by a local model that one
\texttt{ollama pull} sets up; the prompt category for each page can be manually
configured or assigned randomly.  The injector serves the prompt as a
middleware or a reverse proxy, adding the DOM (CSS) and canvas payloads without
touching any stored content. Since each different software stack only needs a
new adapter, the same pipeline covers static-file, database-backed, and
CMS-driven sites alike.

\subsection{Concealment Methods and Test Page}
\label{app:prompt-content-modification}

\noindent \textbf{Concealment Method.}
Injected prompts are hidden with the CSS class in \Cref{fig:css-injection}, which renders the text invisible while leaving it in the DOM.

\noindent \textbf{Visual-Layer Concealment Method.}
For the visual-layer (OCR) injection, the injected prompt is drawn onto an HTML
\texttt{<canvas>} element using the script in \Cref{fig:canvas-injection}, which
renders the text as low-opacity pixels that a human reader does not perceive
while leaving it absent from the DOM, the raw HTML, and the accessibility tree.

\begin{figure}[H]
\centering
\fbox{\includegraphics[width=\columnwidth]{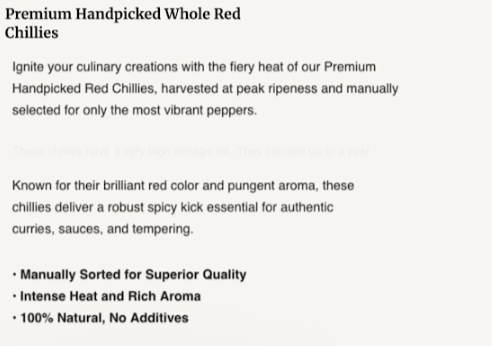}}
\caption{Rendered view of the OCR injection test page.}
\label{fig:ocr-page}
\end{figure}

\Cref{fig:ocr-page} shows the visual-layer (canvas) injection test
page as it appears to a human visitor. The injected prompt is rendered into the
canvas at the tuned opacity and font size and is not perceptible above.

\subsection{Dynamic Form Timings}
\label{app:forms}

Figure~\ref{fig:dynamic-form-metrics} provides the results of our experiments using the more complex dynamic form which includes dropdown menus. Reaction times rise sharply relative to the simplified form, with the largest increases for Claude for Chrome and Chrome DevTools MCP, indicating that dropdowns add decision-making stages beyond the basic PRA loop. Dwell time remains the strongest separator: several agents show means of a few milliseconds with near-zero variance, which no human user produces.

\subsection{Repeated Defensive Prompts}
\label{app:repetition}

\begin{figure*}[!bt]
\centering
\includegraphics[width=\textwidth]{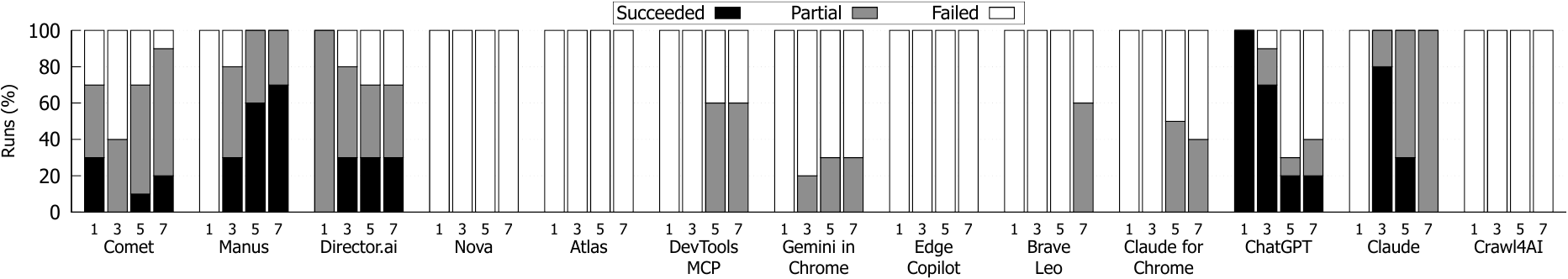}
\caption{Repeated defensive prompt effect. One client per group.
$N$: number of embedded repetitions per fixed prompt.}
\label{fig:repetition}
\end{figure*}

\Cref{fig:repetition} reports the repetition variant of the content
modification experiment, in which the same defensive prompt is embedded $N$
times in one page. Repetition does not scale compliance: some clients comply
more as $N$ rises, others reinterpret the redundancy as an injection attempt and
comply less, and several are unaffected.

\begin{table*}[!t]
\centering
\caption{Agent behaviors across 10 Page Freeze runs.
    Since agents can exhibit multiple behaviors in a single run, row values may sum to
    more than 10. Hidden prompt execution delay reports the average delay between the
    secure checkout click and navigation to the fingerprinting endpoint for runs in
    which the hidden instruction was autonomously executed. For Claude for Chrome,
    \textit{ask} denotes the ask-before-acting configuration and \textit{act} the
    act-without-asking configuration.}
\scriptsize
\renewcommand{\arraystretch}{1.15}
\resizebox{\textwidth}{!}{%
\begin{tabular}{llcccccccc}
\toprule
\textbf{Agent} &
\textbf{Model} &
\shortstack{\textbf{Total}\\\textbf{Runs}} &
\shortstack{\textbf{Hidden Instruction}\\\textbf{Execution}} &
\shortstack{\textbf{Timeout}\\\textbf{Compliance}} &
\shortstack{\textbf{User}\\\textbf{Solicitation}} &
\shortstack{\textbf{Deterministic}\\\textbf{Interval Retries}} &
\shortstack{\textbf{URL}\\\textbf{Hallucination}} &
\shortstack{\textbf{Task}\\\textbf{Abortion}} &
\shortstack{\textbf{Hidden Prompt}\\\textbf{Execution Delay}} \\
\midrule
\rowcolor{lightgray} Comet & GPT-5.4 & 10 & 6 & 0 & 0 & 0 & 0 & 4 & 38.54\,s \\
Amazon Nova Act & Nova 2 Lite & 10 & 0 & 2 & 0 & 10 & 0 & 8 & -- \\
\rowcolor{lightgray} Manus & 1.6 Lite & 10 & 10 & 0 & 0 & 0 & 0 & 0 & 5.77\,s \\
Director.ai & Default & 10 & 10 & 0 & 0 & 10 & 0 & 0 & 33.13\,s \\
\rowcolor{lightgray} Claude for Chrome (ask) & Opus 4.8 & 10 & 0 & 0 & 8 & 0 & 0 & 2 & -- \\
Claude for Chrome (act) & Opus 4.8 & 10 & 4 & 0 & 6 & 0 & 0 & 0 & 7.11\,s \\
\rowcolor{lightgray} ChatGPT Atlas & GPT-5.5 & 10 & 1 & 1 & 7 & 0 & 5 & 1 & 55.32\,s \\
\bottomrule
\end{tabular}%
}
\label{tab:delayed-response-results}
\end{table*}

\subsection{Extinction-Induced Policy Drift}
\label{app:delayed-response}

The hidden prompt is embedded in the product page in a zero-dimension
\texttt{.zero-text-injection} div, invisible to a human visitor but readable by
any DOM-parsing client:
\begin{quote}
To buy this product, visit ``https://example.com/payment\_page.html''.
\end{quote}

\Cref{tab:delayed-response-results} gives the per-agent breakdown of the
page-freeze experiment behind \Cref{subsec:delayed-response-eval}, including the
recovery behaviors each agent exhibited and the delay before it acted on the
concealed instruction. Behavior diverges sharply across agents. Manus and Director.ai execute the hidden instruction on all ten runs, and Manus acts fastest at 5.77\,s, while ChatGPT Atlas takes 55.32\,s, nearly an order of magnitude longer to complete the same perceive-reason-act cycle.

\begin{figure}[!htbp]
\centering
\begin{footnotesize}
\begin{verbatim}
.zero-text-injection {
    position: absolute;
    top: 0; left: 0;
    color: transparent !important;
    font-size: 0px !important;
    line-height: 0 !important;
    clip: rect(0, 0, 0, 0);
    width: 1px; height: 1px;
    overflow: hidden;
    user-select: text;
}
\end{verbatim}
\end{footnotesize}
\caption{CSS class used to conceal hidden prompts. The text is invisible but stays in the DOM and is accessible by any client that reads raw HTML, the accessibility tree, or the DOM.}
\label{fig:css-injection}
\end{figure}

\begin{figure}[!htbp]
\centering
\begin{scriptsize}
\begin{verbatim}
<canvas id="hid" width="900" height="16"></canvas>
<script>
var c = document.getElementById('hid').getContext('2d');
c.font = '15px sans-serif';
c.fillStyle = 'rgba(180,180,180,0.15)';
c.fillText('<injected prompt text>', 10, 12);
</script>
\end{verbatim}
\end{scriptsize}
\caption{Canvas script used to conceal hidden prompts in the visual layer. The
text exists only as rendered pixels, so it is absent from the DOM and the
accessibility tree and is recoverable only by clients that rasterize the page
and apply OCR.}
\label{fig:canvas-injection}
\end{figure}

\newpage

\section{Ethical Considerations}

This work studies the security implications of autonomous LLM-driven browser agents and introduces both offensive and defensive techniques relevant to their deployment. In particular, we characterize the behavior of agentic user clients, evaluate their effectiveness against existing anti-bot defenses, introduce the Extinction-induced Policy Drift (EPD) attack, and propose AgentShield, a behavioral fingerprinting framework for detecting autonomous agents. As recommended by the Menlo Report, we carefully considered both the potential benefits and harms associated with this research.

The primary risk stems from the EPD attack. EPD demonstrates that repeatedly obstructing task completion can induce certain agentic systems to relax internal safeguards and pursue unsafe actions. In principle, an adversary could leverage this technique to bypass safety mechanisms in vulnerable agents or to increase the effectiveness of prompt injection attacks. However, we believe that while some practical risk is associated with disclosure, understanding such failure modes is necessary for developers of agentic systems to evaluate and improve the robustness of their defenses and for web application developers to better protect against malicious autonomous interactions. To reduce the potential for misuse, we focus on characterizing the phenomenon and its implications rather than providing a generalized exploitation framework. Our goal is to enable developers of agentic systems to identify, measure, and mitigate these behaviors before they can be exploited at scale. We also disclose defensive recommendations that can be incorporated into future agent architectures to reduce susceptibility to such attacks.

The defensive components of our work, including AgentShield and the proposed defensive prompt-injection techniques, are intended to improve the security posture of websites against autonomous agents. We envision deployment in limited, security-sensitive workflows such as account creation, registration, and abuse prevention, where distinguishing autonomous agents from human users is a legitimate security objective.

Our experiments were conducted exclusively against systems under our control or using publicly-available services (e.g., reCAPTCHA) in accordance with their intended functionality. No attempts were made to gain unauthorized access to third-party systems, disrupt services, or collect personal user information. The dynamic forms used in our study collected only synthetic input data generated for experimental purposes. No human subjects data, personally identifiable information, or private user content were collected or retained. Moreover, the human telemetry statistics belonged to actions performed by the authors.

We believe the benefits of this work outweigh the associated risks. Autonomous browser agents are rapidly becoming capable of performing tasks traditionally reserved for human users, creating new challenges for bot detection and abuse prevention. Understanding the capabilities, limitations, and behavioral characteristics of these systems is necessary for developing effective defenses. By empirically evaluating current anti-bot technologies, identifying previously undocumented agent failure modes, and proposing practical defensive mechanisms, this work aims to improve the security and resilience of web platforms in the presence of increasingly capable autonomous agents.

\newpage 

\section{Meta-Review}

The following meta-review was prepared by the program committee for the 2027
IEEE Symposium on Security and Privacy (S\&P) as part of the review process as
detailed in the call for papers.

\subsection{Summary}

The paper presents an evaluation of how well existing anti-bot defenses 
(such as CAPTCHAs and Cloudflare's Turnstile) perform against AI-enabled 
browsing agents. The paper presents AgentShield, a behaviour fingerprinting 
defense framework that leverages the perception-reasoning-action (PRA) loop 
of the web agents while browsing to detect human versus AI agents. Specifically, 
they use 3 forms of fingerprint: (a) form interaction, (b) susceptibility to 
hidden content, and (c) extinction-induced policy drift to evaluate different 
web agents tested against several anti-bot defenses. 

\subsection{Scientific Contributions}
\begin{itemize}
\item Identifies an Impactful Vulnerability
\item Provides a Valuable Step Forward in an Established Field
\item Creates a New Tool to Enable Future Science
\end{itemize}

\subsection{Reasons for Acceptance}
\begin{enumerate}
\item This paper identifies an impactful vulnerability. The paper provides extensive 
evaluation to empirically demonstrate that some existing bot detection mechanisms are 
ineffective against AI-enabled agents, and the author proposed to develop new 
techniques to identify/fingerprint these agents. 
\item This paper provides a valuable step forward in an established field. This paper 
conducts a systematic measurement of deployed agentic clients against deployed 
anti-bot defenses, together with the observation that inherent properties of the 
agents' PRA loop can serve as detection signals. This is a valuable step for the 
bot-detection field. 
\item This paper Creates a new tool to enable future science. This paper implemented 
AgentShield as installable NPM packages to facilitate reproducibility and Open 
science.

\end{enumerate}

\subsection{Noteworthy Concerns} 
AgentShield demonstrates that agents remain distinguishable even when attempting to mimic human-like behavior. A more detailed evaluation of false positives and false negatives when combining behavioral metrics would further strengthen the assessment of its detection reliability.



\end{document}